\RequirePackage{amsmath}
\documentclass[10pt]{iopart} 

\usepackage{braket}
\usepackage[dvipdfmx]{graphicx,color}

\usepackage[T1]{fontenc}
\usepackage{graphicx}
\usepackage{bm}
\usepackage{braket}
\usepackage{amssymb}
\usepackage{mathptmx}
\usepackage{xcolor}

\begin{document}

\title[Topical Review: Building large-scale quantum computers with continuous-variable optical technologies
]
{Building a large-scale quantum computer with continuous-variable optical technologies}
\author{Kosuke Fukui \& Shuntaro Takeda}

\address{Department of Applied Physics, School of Engineering, The University of Tokyo, 7-3-1 Hongo, Bunkyo-ku, Tokyo 113-8656, Japan}
\ead{fukuik.opt@gmail.com, takeda@ap.t.u-tokyo.ac.jp}
\vspace{10pt}

\begin{abstract}
Realizing a large-scale quantum computer requires hardware platforms that can simultaneously achieve universality, scalability, and fault tolerance. As a viable pathway to meeting these requirements, quantum computation based on continuous-variable optical systems has recently gained more attention due to its unique advantages and approaches. This review introduces several topics of recent experimental and theoretical progress in the optical continuous-variable quantum computation that we believe are promising. In particular, we focus on scaling-up technologies enabled by time multiplexing, bandwidth broadening, and integrated optics, as well as hardware-efficient and robust bosonic quantum error correction schemes.
\end{abstract}

%
\vspace{2pc}
\ioptwocol

\section{Introduction}
Over the past decades, much effort has been dedicated to developing quantum computers
with the promise of performing previously impossible computing tasks.
Various quantum computing (QC) hardware platforms have emerged,
such as superconducting circuits~\cite{19Arute,21Gong},
trapped ions~\cite{20Pagano,21Pino},
semiconductor quantum dots~\cite{18Watson,21Hendrickx},
and photonic circuits~\cite{20Zhong,21Arrazola}.
One of the leading platforms is the superconducting circuits, as highlighted by a recent demonstration of ``quantum supremacy'' with a 53-qubit superconducting quantum computer~\cite{19Arute}. This device successfully performed a specific computing task that was classically hard to simulate even with the state-of-the-art supercomputer.
The current quantum computers are, however, still too small and noisy
to perform practical quantum algorithms.
Recently, near-term applications with such noisy intermediate-scale quantum computers
have been pursued with several approaches,
such as the development of quantum-classical hybrid algorithms and error mitigation techniques~\cite{20Cerezo,21Endo}.
However, the applicability of these approaches is currently limited and still unclear.
Scaling up quantum computers is an inevitable direction to take advantage of  the full potential of quantum computers.

What are the requirements imposed on hardware platforms to realize practical, large-scale quantum computers? Here we will focus on three requirements. The first requirement is ``universality'': the ability to perform a set of operations necessary to construct arbitrary quantum computation (QC). This requirement is essential to realize general-purpose quantum computers which can run any kinds of quantum algorithms. The second is ``scalability'': the possibility to increase the number of qubits and operational steps efficiently without constraints. The difficulty of this requirement lies in the fact that quantum processors for more and more qubits and operations often suffer from increasing noise channels and more severe restrictions originating from their environmental conditions. The third is ``fault tolerance'': the functionality to correct errors during computation and derive reliable results. Various quantum error-correction (QEC) schemes are already known~\cite{17Campbell}, but they not only require a low error rate below 1{\%} but also incur tremendous resource overhead that makes their implementation a daunting challenge.

As a viable pathway to meeting these severe requirements, QC based on continuous variables (CVs) has recently gained more and more attention.
The mainstream of QC utilizes qubits as fundamental quantum information units and encodes them into two-level physical systems.
In contrast, QC based on CVs is a complementary approach expressing quantum information by quantum states in an infinite-dimensional Hilbert space~\cite{99Lloyd}.
This infinite-dimensional space offers computational models unique to CVs,
such as Gaussian boson sampling and its related algorithms~\cite{17Hamilton,15Huh,18Arrazola,20Banchi},
quantum machine learning based on CVs~\cite{17Lau,schuld2019quantum,19Killoran},
and other CV quantum algorithms and simulations~\cite{15Marshall,18Kalajdzievski,19Arrazora}.
Even more powerful tools in CVs are bosonic quantum error-correcting codes (QECCs)~\cite{01Gottesman,20Terhal}.
The bosonic codes can encode an error-correctable logical qubit into a single-mode quantum state by exploiting its infinite-dimensional space as a resource to introduce redundancy.
These codes are more efficient than the standard error-correcting codes requiring additional two-level physical systems
to increase the dimensionality and introduce redundancy.
Therefore, the bosonic codes potentially reduce resource overhead and thus provide hardware-efficient implementations of QEC.

Several quantum systems that can potentially perform CVQC are known,
such as microwave modes in superconducting circuits{~\cite{blais2021circuit,hillmann2020universal}, vibrational modes of trapped ions~\cite{chen2021quantum,ortiz2017continuous} 
or mechanical oscillators~\cite{aspelmeyer2014cavity},
atomic spin ensembles~\cite{hammerer2010quantum,pezze2018quantum}, and optical modes~\cite{15Andersen,19Pfister,19Takeda2}. 
For microwave and optical modes, several architectures for fault-tolerant CVQC have been recently proposed
~\cite{20Terhal,noh2020fault,grimsmo2021quantum,21Bourassa,21Larsen,tzitrin2021fault}.
Here in this review, we shed light on ``optical'' CVQC,
expecting its potential to satisfy the universality, scalability, and fault tolerance simultaneously.
The optical CV approach has recently made remarkable progress in realizing scalability, as exemplified by
the deterministic generation of large-scale entangled states~\cite{14Chen,19Asavanant,19Larsen} and their application to many steps of QC~\cite{20Asavanant,20Larsen}.
This scalability comes from deterministic quantum resources available in CV schemes
and optical multiplexing techniques to increase optical modes efficiently.
In addition, optical platforms suffer from fewer limitations to scalability since
they do not require severe environments such as very low temperature or vacuum.
In terms of fault tolerance, the optical CV approach has unique advantages.
Optical quantum systems are inherently robust because light has almost no interaction with environments.
The dominant cause of errors is optical losses, but the bosonic QECCs
can deal with such loss errors in the hardware-efficient ways.
The last requirement, universality, has long been one of the major difficulties in the optical QC.
This is because the strong optical nonlinearity required for universal quantum operations has been hard to achieve deterministically.
However, recent progress in the CV approach has found several feasible methodologies to
realize deterministic nonlinear optical quantum operations~\cite{11Marek,16Miyata,18Marek}.
For these reasons, various types of optical architectures for scalable, fault-tolerant, and universal CVQC have recently been proposed~\cite{21Bourassa,20Fukui,21Larsen}.

This review aims to introduce several topics of recent experimental and theoretical progress in the optical CVQC that we believe opens a promising path to realizing universality, scalability, and fault tolerance at the same time.
First, Section II focuses on the universality.
We define basic operations in the CVQC
and show how to perform the universal QC in theory.
We also outline the one-way CVQC model,
which is the leading approach in this field.
Then, Section III focuses on the scalability.
After briefly summarizing elementary technologies developed thus far,
we mainly concentrate on experimental progress in three topics related to scalability,
including optical multiplexing, bandwidth broadening, and integrated photonic chips.
Section IV focuses on the fault tolerance.
We introduce bosonic QECCs, focusing on Gottesman-Kitaev-Preskill (GKP) qubits~\cite{01Gottesman} in particular.
We also discuss the methodologies for the fault-tolerant optical CVQC with GKP qubits.
Finally, Sec. V summarizes and concludes this review.

\section{Universal quantum computation with CVs}
\subsection{Continuous variables} 
In the beginning, we review the operators and the measurement of quantum states for CVQC.
The Hamiltonian of a quantized electromagnetic mode is described by that of a quantum harmonic oscillator, where quadratures of a single mode $k$ correspond to the harmonic oscillator's position and momentum operators $\hat{q}_{k}$ and $\hat{p}_{k}$, respectively.
The Hamiltonian of the quantum harmonic oscillator can be written by these operators as
\begin{equation}
\hat{H}_k=\frac{1}{2}(\hat{p}^2_{k}+\omega^2_{k}\hat{q}^2_{k})=\hbar \omega_k \left(\hat{a}^{\dag}_{k} \hat{a}_{k}+\frac{1}{2}\right),
\end{equation}
where $\hat{a}_{k}$ and $\hat{a}^{\dag}_{k}$ are annihilation and creation operators, $\omega_k$ is the center frequency $\omega$ of the mode $k$, and $\hbar$ is the Planck constant divided by $2\pi$.
The position and momentum operators are described by creation and annihilation operators as
\begin{equation}
\hat{q}_{k}=\sqrt{\frac{\hbar}{2\omega}_{k}}(\hat{a}_{k}+\hat{a}^{\dag}_{k}), \hspace{10pt}
\hat{p}_{k}=-i\sqrt{\frac{\hbar {\omega}_{k}}{2}}(\hat{a}_{k}-\hat{a}^{\dag}_{k}),
\end{equation}
which satisfy the commutation relation, $[\hat{q}_k,\hat{p}_{k'}]=i\hbar \delta_{kk'}$, $[\hat{a}_k, \hat{a}^{\dag}_{k'}] = \delta_{kk'}$, and $[\hat{a}_k,\hat{a}_{k'}]=0$ ($\delta_{kk'}$ is the Kronecker delta).
The position and momentum quadratures then represent a conjugate pair, and the well-known uncertainty relation becomes
\begin{equation}
\langle(\Delta \hat{q}_k)^2\rangle\langle(\Delta \hat{p}_k)^2\rangle\geq\frac{1}{4} |\langle \left[\hat{q}_k,\hat{p}_k \right]\rangle|^2=\frac{\hbar^2}{4}.
\end{equation}
In the following, we will use the position and momentum quadratures with ${\omega}_{k}=1$ and $\hbar=1$, and may omit the label $k$.

\begin{table*}
\caption{Comparison between qubits and qumodes in quantum information processing.}
\label{comparison}
\renewcommand{\arraystretch}{1.2}
\begin{tabular}{|c|p{6cm}|p{7cm}|}
\hline
& \hspace{40pt}Discrete variables (qubits) & \hspace{40pt}Continuous variables (qumodes) \\\hline
\hline
Computational basis & $\{\ket{0}_{\rm L},\ket{1}_{\rm L}\}$ & $\{\ket{s}_q\}_{s\in \mathbb{R}}$ \\ \hline
Conjugate basis & $\{\ket{\pm}_{\rm L}=\frac{1}{\sqrt{2}}(\ket{0}_{\rm L}\pm\ket{1}_{\rm L})\}$ &$ \{ \ket{t}_p=\frac{1}{\sqrt{2\pi}}\int_{-\infty}^{\infty}ds\hspace{2pt}  e^{ist}\ket{s}_q \}_{t\in \mathbb{R}} $\\ \hline
Encoding & $\ket{\psi}=\alpha\ket{0}_{\rm L}+\beta\ket{1}_{\rm L}$\hspace{5pt} $(|\alpha|^2+|\beta|^2=1)$ & $\ket{\psi}=\int_{-\infty}^{\infty}ds\hspace{0pt}\psi(s)\ket{s}_q$ $\hspace{2pt}$ $(\int_{-\infty}^{\infty}ds |\psi(s)|^2=1)$\\ \hline
Detector & Photon detector & Homodyne detector \\ \hline
&  \hspace{3pt}Bit-flip gate: $\hat{X}$ &\hspace{3pt}Displacement in $\hat{q}$ basis: $\hat{X}(v)=e^{-iv\hat{p}}$ $(v \in \mathbb{R})$\\
 &\hspace{7pt}$\hat{X}\ket{0}_{\rm L}=\ket{1}_{\rm L}, \hspace{2pt}\hat{X}\ket{1}_{\rm L}=\ket{0}_{\rm L}$ & \hspace{8pt}$\hat{X}(v)\ket{s}_q=\ket{s+v}_q$ \\
 &\hspace{3pt}Phase-flip gate: $\hat{Z}$ &\hspace{3pt}Displacement in $\hat{p}$ basis :$\hat{Z}(u)=e^{iu\hat{q}}$ $(u \in \mathbb{R})$\\
&\hspace{7pt}$ \hat{Z}\ket{0}_{\rm L}=\ket{0}_{\rm L}, \hspace{2pt}\hat{Z}\ket{1}_{\rm L}=-\ket{1}_{\rm L}$ &  \hspace{8pt}$\hat{Z}(u)\ket{t}_p=\ket{t+u}_p$ \\ 
Quantum gate   &  \hspace{3pt}Hadamard gate: $\hat{H}$ &\hspace{1pt} Fourier gate: $\hat{R}{(\frac{\pi}{2})}=e^{i\frac{\pi}{4} (\hat{q}^2+\hat{p}^2)}$ \\
&\hspace{7pt}$ \hat{H}\ket{0}_{\rm L}=\ket{+}_{\rm L}, \hspace{2pt}\hat{H}\ket{1}_{\rm L}=\ket{-}_{\rm L}$ &\hspace{5pt}{ $\hat{R}{(\frac{\pi}{2})}\ket{s}_q=\ket{s}_p$,\hspace{7pt}$\hat{R}{(\frac{\pi}{2})}\ket{t}_p=\ket{-t}_q$}   \\
&\hspace{3pt}Controlled-NOT (CX) gate: $\hat{CX}$   &\hspace{3pt}CX gate: $\hat{CX} =e^{-i \hat{q}_{1} \hat{p}_{2}}$ \\
&\hspace{7pt}$\hat{CX}\ket{0}_{\rm L}\ket{0(1)}_{\rm L}=\ket{0}_{\rm L}\ket{0(1)}_{\rm L}$ &\hspace{8pt}$\hat{CX}\ket{s_1}_{q_1}\ket{s_2}_{q_2}=\hat{CX}\ket{s_1}_{q_1}\ket{s_2+s_1}_{q_2}$     \\  
&\hspace{7pt}$\hat{CX}\ket{1}_{\rm L}\ket{0(1)}_{\rm L}=\ket{1}_{\rm L}\ket{1(0)}_{\rm L}$ &\hspace{8pt}$\hat{CX}\ket{t_1}_{p_1}\ket{t_2}_{p_2}=\hat{CX}\ket{t_1-t_2}_{p_1}\ket{t_2}_{p_2}$     \\     \hline
Carrier  &  Degrees of freedom of a photon & Quadratures of a light field \\ \hline 
\end{tabular}
\renewcommand{\arraystretch}{1.0}
\end{table*}

The measurement in the quadrature eigenbasis is implemented by the homodyne measurement, which is an essential part of implementing CVQC. 
In the (balanced) homodyne measurement~\cite{yuen1983noise}, the signal mode to be measured and the local oscillator (LO) mode are coupled by a beam splitter with a transmissivity 50\%, and these two modes are measured by photodetectors. 
The difference of the output photocurrents is given by $\delta \hat{i}=c(\hat{a}_{\rm LO}^{\dag}\hat{a}_{\rm sig}+\hat{a}_{\rm LO}\hat{a}^{\dag}_{\rm sig})$ with constant $c$, where $\hat{a}_{\rm sig}(\hat{a}_{\rm LO})$ is the annihilation operator for the signal (LO) mode. Assuming a large amplitude of the LO mode, i.e., $\hat{a}_{\rm LO} \to \alpha_{\rm LO}=|\alpha_{\rm LO}|e^{i\Phi}$ with the relative phase between signal and LO modes $\Phi$, we can describe $\delta \hat{i}$ as $\delta \hat{i}\propto|\alpha_{\rm LO}|(\hat{q}_{\rm sig}{\rm cos}\Phi + \hat{p}_{\rm sig}{\rm sin}\Phi) $.
By changing $\Phi$ we can measure the quadratures of the signal mode. 
The cases for $\Phi=0$ and $\pi/2$ correspond to the measurement in the $q$ and $p$ quadratures, respectively.

\subsection{Quantum computation with CVs} \label{subsec:cvqc}
\subsubsection{Continuous and discrete variables} 
For quantum information processing based on the qubit, the eigenbasis $\left\{\ket{0}_{\rm L},\ket{1}_{\rm L} \right\}$ is used for the computational basis. 
For CVs, the eigenbasis of $\hat{q}$, $\{ \ket{s}_q\}_{s\in \mathbb{R}}$, is conventionally used as the computational basis, and it spans an infinite-dimensional Hilbert space.
Comparison between qubits and qumodes in quantum information processing is summarized in Table~\ref{comparison}, where qumodes are CV equivalent of the qubits.
Reviews for quantum information processing with CVs were provided in Refs.~\cite{braunstein2005quantum,weedbrook2012gaussian,19Pfister,19Takeda2}.

QC has a great deal of potential to efficiently solve some hard problems for conventional computers.
On the other hand, the Gottesman-Knill theorem for qubits~\cite{gottesman1998heisenberg} shows that a class of QC with qubits which employs only Clifford gates as well as projective measurements in the computational basis can be efficiently simulated by a classical computer.
Thus, non-Clifford gates play an important role in providing a speedup over a classical computer, and allow us to realize universality defined by the condition that any unitary transformation can be implemented with arbitrary accuracy.
It is known that universality can be achieved by a finite set of Clifford and non-Clifford gates, which is called a universal gate set.

CVQC was proposed by Lloyd and Braunstein~\cite{99Lloyd}, showing the way to use CVs for QC and the universal gate set for CVQC.
For universal CVQC, non-Gaussian gates corresponding to the non-Clifford gates are required.
The CV analogue of the Gottesman--Knill theorem for qubits~\cite{gottesman1998heisenberg} was formulated by Bartlett {\it et al.}~\cite{02Bartlett}, showing that CVQC cannot overcome a classical computer if initial states, gates, and measurements are all Gaussian.
We note that Gaussian (non-Gaussian) gates linearly (nonlinearly) transform quadratures by Hamiltonians with the second or lower order (the third or higher order) of quadrature operators, requiring only second-order (third or higher-order) optical nonlinearity.
In the following Secs. 2.2.2 and 2.2.3, we see the universal gate set consisting of a finite set of Gaussian and non-Gaussian gates.

\subsubsection{Gaussian quantum gates}
A typical Gaussian quantum gate set for universality consists of \{$\hat{Z}(t)$, $\hat{P}(\eta)$, $\hat{R}{(\theta)}$, $\hat{CZ}$\}, where $\hat{Z}(t)$ is $p$ quadrature displacement, $\hat{P}(\eta)$ is shearing (also referred to as the phase gate), $\hat{R}{(\theta)}$ is rotation (also referred to as the phase shift), and $\hat{CZ}$ is the controlled-phase (CZ) gate.
The above Gaussian gates can be partly substituted by $q$ quadrature displacement $\hat{X}(t)$, squeezing $\hat{S}(r)$, the controlled-NOT (CX) gate, and a beam-splitter coupling $\hat{BS}{(\theta)}$.
In terms of an experimental realization in an optical setup, the Gaussian quantum gate set for universality, \{$\hat{S}(r)$, $\hat{Z}(t)$ (or $\hat{X}(t)$), $\hat{R}{(\theta)}$, $\hat{BS}{(\theta)}$\}, is also employed since these Gaussian gates are more feasible than others. 

We see the Gaussian quantum gates in more detail.
Displacement operators in position and momentum quadratures are defined as $\hat{X}(v)=e^{-iv\hat{p}}$ and $\hat{Z}(u)=e^{iu\hat{q}}$ with $v, u \in \mathbb{R}$, respectively.
These operators act on the quadrature eigenbasis as
\begin{equation}
\hat{X}(v)\ket{s}_q=\ket{s+v}_q, \hat{Z}(u)\ket{t}_p=\ket{t+u}_p.
\end{equation}
In the Heisenberg picture, $\hat{X}(v)$ and $\hat{Z}(u)$ transform the position and momentum quadratures as $ \hat{q} \to \hat{q}+v$ and $ \hat{p} \to \hat{p}+u$, respectively.  
$\hat{X}(v)$ and $\hat{Z}(u)$ are analogous to bit- and phase-flip gates for the qubit, called Pauli $X$ and $Z$ gates, which act on the computational basis as $\hat{X}\ket{0}_{\rm L}(\hat{X}\ket{1}_{\rm L})=\ket{1}_{\rm L}(\ket{0}_{\rm L})$ and $\hat{Z}\ket{0}_{\rm L}(\hat{Z}\ket{1}_{\rm L})=\ket{0}_{\rm L}(-\ket{1}_{\rm L})$, respectively.
Additionally, $\hat{X}(v)$ and $\hat{Z}(u)$ are generalized to the Weyl-Heisenberg group of the displacement operator $\hat{D}(\alpha)$ which is any linear combination of $\hat{X}(v)$ and $\hat{Z}(u)$ and defined as $\hat{D}(\alpha)={\rm exp}(\alpha \hat{a}^{\dag}-\alpha^{\ast}\hat{a})$ with the complex amplitude $\alpha=(v+iu)/\sqrt{2}$.

The squeezing, phase, and rotation gates are defied as $\hat{S}(r)=e^{i \frac{r}{2} (\hat{q}\hat{p}+\hat{p}\hat{q})}$, $\hat{P}(\eta)=e^{i\frac{\eta}{2} \hat{q}^2}$, and $\hat{R}{(\theta)}=e^{i\frac{\theta}{2} (\hat{q}^2+\hat{p}^2)}$, respectively. In the Heisenberg picture, they transform quadratures as
\begin{eqnarray}
&\hat{S}(r)&:\hat{q}\to e^{-r}\hat{q}, \hspace{10pt} \hat{p}\to e^{r}\hat{p},\\
&\hat{P}(\eta) &: \hat{q}\to\hat{q}, \hspace{10pt} \hat{p}\to\hat{p}+\eta\hat{q}, \\
&\hat{R}{(\theta)}&:\hat{q}\to\hat{q}\mathrm{cos}\theta -\hat{p}\mathrm{sin}\theta, \hspace{10pt} \hat{p}\to \hat{q}\mathrm{sin}\theta+\hat{p}\mathrm{cos}\theta. \label{rotation}
\end{eqnarray}

In addition to the single-mode Gaussian gates described above, two-mode Gaussian gates are also required for CVQC. The typical example is the CZ gate $\hat{CZ}=e^{i \hat{q}_{1}\hat{q}_{2}}$, which transforms the quadratures for the modes 1 and 2 as 
\begin{equation}
\hat{q}_{1}\to \hat{q}_{1}, \hspace{10pt} \hat{p}_{1}\to \hat{p}_{1}+\hat{q}_{2}, \hspace{10pt}
\hat{q}_{2} \to \hat{q}_{2}, \hspace{10pt} \hat{p}_{2} \to \hat{p}_{2}+\hat{q}_{1}.
\end{equation}
In addition, the CX gate is defined as $\hat{CX} =e^{-i \hat{q}_{1}\hat{p}_{2}}$, which  transforms the quadratures as 
\begin{equation}
\hat{q}_{1}\to \hat{q}_{1}, \hspace{10pt} \hat{p}_{1}\to \hat{p}_{1}-\hat{p}_{2}, \hspace{10pt}
\hat{q}_{2} \to \hat{q}_{2}+\hat{q}_{1}, \hspace{10pt} \hat{p}_{2} \to \hat{p}_{2}.
\end{equation}
The CX gate for CVs is also referred to as the quantum-nondemolition (QND) gate or the sum gate.
We note that, more generally, CZ and CX gates for CVs can be defined as $\hat{CZ}(g) =e^{ig \hat{q}_{1}\hat{q}_{2}}$ and $\hat{CX}(g) =e^{-ig \hat{q}_{1}\hat{p}_{2}}$, respectively, with interaction strength $g$, but $g$=1 is used throughout this review.

Experimentally, the CZ and CX gates can be implemented by linear optics and ancillary squeezed vacuum states, $\hat{S}(r)\ket{0}$~\cite{05Filip,08Yoshikawa,18Shiozawa}.
However, the ancilla state introduces additional errors due to finite squeezing.
Instead, a beam splitter operation can be generally used for a two-mode gate due to the experimental feasibility in an optical setup.
The beam-splitter operation $\hat{BS}{(\theta)}$ with the parameter $\theta$ is defined as $\hat{BS}{(\theta)}=e^{\frac{\theta}{2} (\hat{a}_1^{\dag}\hat{a}_2-\hat{a}_1\hat{a}_2^{\dag})}=e^{\frac{i\theta}{2} (\hat{q}_1\hat{p}_2+\hat{p}_1\hat{q}_2)}$,
where $\hat{a}_{1(2)}$ is the annihilation operator for the mode {1(2)} and $\theta$ determines the transmissivity of the beam splitter $T={\rm cos}^2\frac{\theta}{2} \in[0,1]$. 
$\hat{BS}{(\theta)}$ transforms the quadratures for two modes 1 and 2 as 
\begin{eqnarray}
\hat{q}_{1} &\to& \sqrt{T}\hat{q}_{1}+\sqrt{R}\hat{q}_{2}, \hspace{12pt} 
\hat{p}_{1} \to \sqrt{T}\hat{p}_{1}+\sqrt{R}\hat{p}_{2},\\
\hat{q}_{2} &\to& -\sqrt{R}\hat{q}_{1}+\sqrt{T}\hat{q}_{2}, \hspace{5pt} 
\hat{p}_{2} \to -\sqrt{R}\hat{p}_{1}+\sqrt{T}\hat{p}_{2},
\end{eqnarray}
where $R=1-T$ is the reflectivity of the beam splitter. 

Note that a combination of the above Gaussian gates can generate any Gaussian gate which transforms the quadratures linearly in the Heisenberg picture. On the other hand, these gates provide only Gaussian operations, and are thus not sufficient to generate an arbitrary unitary gate.

\subsubsection{Non-Gaussian quantum gates and universality}\label{subsubsec:nongauss}
Next, we see the non-Gaussian quantum gates for universality. 
A typical non-Gaussian gate is a unitary $\hat{U}_{n}(t)=e^{i\frac{t}{n}\hat{q}^{n}}$ with a natural number $n \geq 3$ and the nonlinear interaction strength $t$.
The unitary $\hat{U}_{3}(t)$ is referred to as the cubic phase gate.
We can see $\hat{U}_{1}(t)$ and $\hat{U}_{2}(t)$ correspond to $\hat{Z}(t)$ and $\hat{P}(t)$, respectively.
In the Heisenberg picture, $e^{i\frac{t}{n}\hat{q}^{n}}$ with $n\geq3$ nonlinearly transforms quadratures as 
\begin{equation}
\hat{q}\to\hat{q}, \hspace{10pt} \hat{p}\to\hat{p}+t\hat{q}^{n-1}.
\end{equation}
It is known that an arbitrary unitary operator can be constructed from Gaussian gates and $\hat{U}_{3}(t)$.
Thus, the universal gate set for an optical setup is given by a finite set of gates, e.g.,\{$\hat{S}(r)$, $\hat{Z}(t)$ (or $\hat{X}(t)$), $\hat{R}{(\theta)}$, $\hat{BS}{(\theta)}$,$\hat{U}_{3}(t)$\}.

The way to construct approximately an arbitrary unitary operator was introduced by Lloyd {\it et al.}~\cite{99Lloyd} and formulated by Sefi {\it et al.}~\cite{sefi2011decompose,sefi2013measurement}. 
Also, there have been many studies on the decomposition of an arbitrary or specific unitary operator exactly or approximately~\cite{kalajdzievski2018continuous,kalajdzievski2019exact,douce2019probabilistic,
annabestani2020towards,kalajdzievski2021exact}.
An arbitrary unitary $\hat{U}$ is described as $\hat{U}={\rm exp}(i\hat{H})$, and $\hat{H}$ is the Hermitian operator consisting of a sum of products of quadrature operators.
Universality is realized when $\hat{H}$ can be approximated with arbitrary precision by using Hamiltonians composing the universal gate set.
To decompose and approximate a given Hamiltonian, 
the following approximations for Hamiltonian operators ${\hat{A}}$ and ${\hat{B}}$ are typically used:
\begin{eqnarray}
e^{it(\hat{A}+\hat{B})} &\approx&e^{i\hat{A}\frac{t}{2}}e^{i\hat{B}t}e^{i\hat{A}\frac{t}{2}}+ O(t^3),\label{decomp1}\\
e^{t^2[\hat{A},\hat{B}]} &\approx&e^{i\hat{B}t}e^{i\hat{A}t}e^{-i\hat{B}t}e^{-i\hat{A}t}+ O(t^3),\label{decomp2}
\end{eqnarray}
where $t$ is a time to apply each of the Hamiltonians~\cite{suzuki1990fractal,sefi2011decompose}. 
When we apply these operators repeatedly by replacing $t$ with $t/N$, Eqs.~(\ref{decomp1}) and~(\ref{decomp2}) are transformed as 
\begin{eqnarray}
e^{it(\hat{A}+\hat{B})} &\approx&(e^{i\hat{A}\frac{t}{2N}}e^{i\hat{B}\frac{t}{N}}e^{i\hat{A}\frac{t}{2N}})^N+ O\left( \frac{t^3}{N^2}\right),\\
e^{t^2[\hat{A},\hat{B}]} &\approx&(e^{i\hat{B}\frac{t}{N}}e^{i\hat{A}\frac{t}{N}}e^{-i\hat{B}\frac{t}{N}}e^{-i\hat{A}\frac{t}{N}})^{N^2}+ O\left( \frac{t^3}{N}\right),
\end{eqnarray}
where $N$ is the number of repetitions of sequential operations consisting of operators $\hat{A}$ and $\hat{B}$.
Thus, we can realize $(\hat{A}+\hat{B})$ and $[\hat{A},\hat{B}]$ from $\hat{A}$ and $\hat{B}$ with arbitrary precision in principle, assuming sufficiently large $N$ so that $O\left( {t^3}/{N^2}\right),O\left( {t^3}/{N}\right)\mapsto0$.
This fact enables us to construct higher-order Hamiltonians of $\hat{q}$ and $\hat{p}$ from lower-order ones. Additionally, the several exact gate decompositions without any approximations have been shown~\cite{sefi2011decompose,kalajdzievski2021exact}, e.g.,
\begin{eqnarray}
\hat{q}^{m+1}&=&-\frac{2}{3m}[\hat{q}^m,[\hat{q}^3,\hat{p}^2]],\\
\hat{q}^{m}p^{n}+\hat{p}^{n}\hat{q}^{m}&=&-\frac{4i}{(n+1)(m+1)}[\hat{q}^{m+1},\hat{p}^{n+1}] \nonumber \\
&-&\frac{1}{n+1}\sum_{k=1}^{n-1}[\hat{p}^{n-k},[\hat{q}^m,\hat{p}^k]].
\end{eqnarray}
The above results allow us to implement an arbitrary Hamiltonian and thus an arbitrary unitary from a universal gate set composed of a finite number of gates.

\begin{figure}[t]
 \centering \includegraphics[angle=0, scale=1.0]{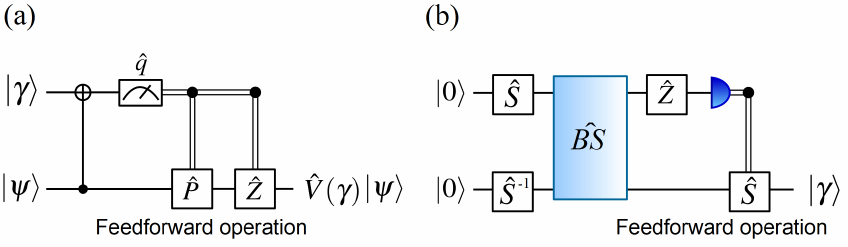} 
\caption{Cubic phase gate. (a)The cubic phase gate by the measurement-induced scheme using the cubic phase state. (b) The preparation of the cubic phase state~\cite{09Gu,sabapathy2018states}.}
\label{cpg}
\end{figure}
\subsubsection{Non-Gaussian gates in an optical setup}\label{subsubsec:cpgexperimental}
As already mentioned in the previous section, universal QC can be implemented by a finite set of Gaussian gates and a single non-Gaussian gate. Optical non-Gaussian gates require third or higher-order optical nonlinear effects. However, such nonlinearity is extremely weak in general and difficult to achieve for optical quantum states at the single-photon level. In fact, this lack of sufficient nonlinearity has been a major problem in realizing two-qubit gates in the optical discrete variable approach. One possible way to circumvent this problem is to introduce ancillary photons and projective measurement to probabilistically induce effective nonlinearity, as proposed by Knill, Laflamme, and Milburn~\cite{knill2001scheme}.
However, this scheme requires a substantial resource overhead to make the two-qubit gate deterministic. Another way is to enhance the nonlinearity by using strong interaction between Rydberg atoms~\cite{tiarks2019photon},
or by coupling two-level systems with photons in cavities or waveguides~\cite{hacker2016photon,chen2021two,zhou2017single,chen2017exact,chen2018entanglement}.

In the case of the CV approach, a simple measurement-based scheme to deterministically perform a non-Gaussian gate was proposed by Gottesman, Kitaev, and Preskill. Figure 1(a) shows their scheme to perform the cubic phase gate $\hat{V}(\gamma)=\hat{U}_{3}(\gamma)={\rm exp}(i\frac{\gamma}{3}\hat{q}^3)$ with $\gamma \in \mathbb{R}$.
In their scheme, the cubic phase state is used as a resource to implement the cubic phase gate, where the ideal cubic phase state is defined as $\ket{\gamma}=\hat{V}(\gamma)\ket{0}_p\propto\int ds e^{i\frac{\gamma}{3} s^3}\ket{s}_q$.
Figure~\ref{cpg}(a) shows the implementation of the cubic phase gate on the input state $\ket{\psi}$. 
In Fig.~\ref{cpg}(a), the CX gate between $\ket{\gamma}$ and $\ket{\psi}$ is implemented, and the upper mode is measured in the $q$ quadrature.
After the measurement, the $p$ quadrature for the input state is transformed as $\hat{p}\to\hat{p}+\gamma \hat{q}^2+2m\gamma\hat{q}+m^2\gamma$ in the Heisenberg picture, where $m$ is the homodyne measurement outcome. 
Finally, $\hat{V}(\gamma)\ket{\psi}$ is obtained after canceling the terms $2m\gamma \hat{q}$ and $m^2\gamma$ via feedforward operations $\hat{P}(-2m\gamma)$ and $\hat{Z}(-m^2\gamma)$ depending on $m$.

The state $\ket{\gamma}$ can be prepared approximately by using squeezed vacuum states, a beam-splitter coupling, a displacement operation, and an adaptive squeezing operation depending on the result of a photon counting measurement~\cite{sabapathy2018states}, as shown in Fig.~\ref{cpg}(b).
This scheme, however, requires a high squeezing level for the preparation of the cubic phase gate. 
For the experimental feasibility, thus, a superposition of Fock states can be alternatively used to approximate the cubic phase state. For example, the state $\ket{\phi}_{\rm cps}\propto \ket{0}+c\ket{3}$ with $c={i\gamma \sqrt{3}}/{2}$ has the position wave function $\phi(q)\propto 1+i\gamma(q^3-3q/2)$.
Since the wave function of $\ket{\gamma}$ can be described as $1+i\gamma q^3 +O(\gamma^2)$, the state $\ket{\phi}_{\rm cps}$ effectively approximates $\ket{\gamma}$ with $\gamma \ll 1$ except for the term $3\gamma q/2$ whose effect can be canceled out later.
For the implementation of a cubic phase gate including the preparation of the cubic phase state, there are many theoretical efforts~\cite{16Miyata,arzani2017polynomial,sabapathy2018states,18Marek,sabapathy2019production} and experimental efforts described in Sec.~\ref{subsubsec:building_blocks}.

\subsection{One-way CVQC}
 
\subsubsection{One-way QC}
A model of QC analogous to classical computation is called a quantum circuit model. In the circuit model, unitary quantum gates are performed on the input quantum states in a quantum register. 
After the unitary time evolution required for the algorithm, the quantum states are measured in the register.
To implement a practical quantum algorithm, this model requires  a number of deterministic unitary quantum gates with coherent control of individual qubits~\cite{10.5555/1972505}. 

Instead of the quantum circuit model, Raussendorf and Briegel formulated one-way QC~\cite{raussendorf2001one}, also called measurement-based QC.
Qubit-based one-way QC provides the ability to perform universal QC using only measurements of the qubits composing the multipartite entangled state, called the cluster state; accordingly, an arbitrary quantum algorithm can be performed via a sequence of single-qubit measurements on qubit-based cluster states.
In an optical setup, the measurement of qubits is usually easier than the quantum gate based on the quantum circuit model, which requires coherent control of qubits.
Thus, one-way QC has been a promising model to implement large-scale QC in optical systems if one can prepare a large-scale cluster state.
Later, one-way CVQC was established by Menicucci {\it et al.}~\cite{06Menicucci} and Gu {\it et al.}~\cite{09Gu}.
The CV version of the cluster state can implement any computational algorithm based on CVQC.

The cluster state is typically described by a graph $G=(V,E)$ with the sets of vertices $V$ and edges $E$, as illustrated in Fig.~\ref{cluster}(a), where the elements of $V$ and $E$ correspond to qubits (qumodes) and CZ gates, respectively. 
The cluster state for qubits is generated by implementing CZ gates between neighboring qubits which are initially prepared to $\ket{+}_{\rm L}=(\ket{0}_{\rm L}+\ket{1}_{\rm L})/\sqrt{2}$, as illustrated in Fig.~\ref{cluster}(b).
The CV cluster state is generated by performing the CZ gate between neighboring qumodes which are initially prepared to momentum-squeezed vacuum states, as illustrated in Fig.~\ref{cluster}(c).
The cluster state is efficiently described by the stabilizer formalism. 
For CVs, the operator $\hat{X}(v)$ is a stabilizer operator for the state $\ket{0}_{p}$ because of $\hat{X}(v)\ket{0}_p=\ket{0}_p$.
For the CV cluster state, the stabilizers for $n$ modes can be defined as the set of $\hat{\mathfrak{S}}_i(v)=\hat{X}_i(v) \prod_{j\in N(i)} \hat{Z}_j(v)$ for $i=1,\cdots, n$ and for all $v\in \mathbb{R}$, where $N(i)$ means the set of vertices in the neighborhood of the vertice $i$.
Here $\hat{\mathfrak{S}}_i(v)$ is described by using nullifiers $\hat{H}_i$ as $\hat{\mathfrak{S}}_i(v)=e^{-iv{\hat H}_i}$ for all $v\in \mathbb{R}$. 
Considering the expression of $\hat{\mathfrak{S}}_i(v)$ for the CV cluster state, $\hat{H}_i$ can be written as $\hat{p}_i-\sum_{j\in N(i)}\hat{q}_j$.
For a simple example, the nullifiers of the cluster state in Fig.~\ref{cluster}(d) are given by $\hat{p}_1-\hat{q}_2,$ $\hat{p}_2-\hat{q}_1-\hat{q}_3,$ and $\hat{p}_3-\hat{q}_2$. 
The stabilizer generators are written as $\hat{X}_1(v)\hat{Z}_2(v)$, $\hat{Z}_1(v)\hat{X}_2(v)\hat{Z}_3(v)$, and $\hat{Z}_2(v)\hat{X}_3(v)$ for all $v$. 
The +1 eigenstate of the stabilizer generators, however, requires infinite squeezing and thus infinite energy.
Since such a state is not realistic, a common way to approximate the CV cluster state is to replace each eigenstate $\ket{0}_p$ with a squeezed vacuum state whose variance in the $p$ quadrature is equal to $ \langle(\Delta \hat{p})^2\rangle =\frac{1}{2}e^{-2r}$ with a finite squeezing parameter $r$.

Universal one-way CVQC requires the preparation of two-dimensional cluster states where qumodes are connected in a two-dimensional array. The preparation of three-dimensional cluster states has also been considered to introduce fault tolerance into one-way QC~\cite{raussendorf2007fault,raussendorf2007topological}.
To prepare the cluster state in an optical setup, a beam-splitter operation is generally used for a two-mode gate, instead of the CZ gate which requires a complicated setup and introduces additional noise from finitely-squeezed ancilla states~\cite{05Filip,08Yoshikawa,18Shiozawa}.
Thus, the CZ gates are replaced by beam-splitter networks with appropriate configuration and parameters. The way to generate the cluster state using beam-splitter networks has been developed over the past two decades~\cite{zhang2006continuous,07vanLoock,11Menicucci,13Yokoyama,
16Yoshikawa,19Asavanant,19Larsen,20Wu,20Fukui}, and related experimental progress is explained in Sec.~\ref{subsec:multiplexing}.

\begin{figure}[t]
 \centering \includegraphics[angle=0, scale=0.95]{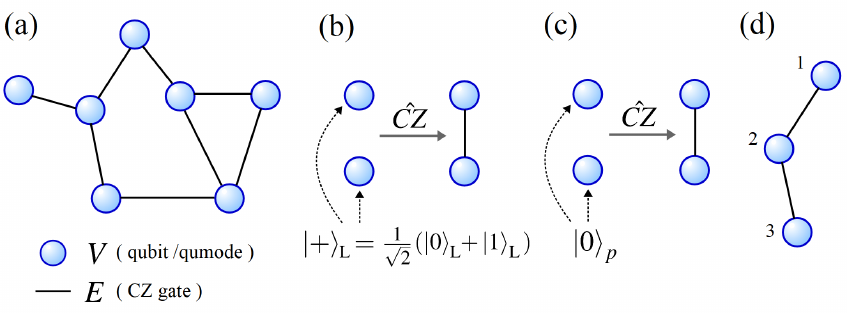} 
\caption{Cluster state. (a) Graph $G=(V,E)$  (b) The cluster state for qubits. (c) The CV cluster state. (d) The three mode cluster state.}
\label{cluster}
\end{figure}

\subsubsection{Quantum gates in one-way CVQC}\label{subsubsec:onewaygates}
 One-way CVQC is performed on the CV cluster state described above.
As a simple example, we consider the Fourier gate using the entangled pair prepared by the CZ gate. 
Figures~\ref{mbqcgate}(a) and (b) show the Fourier gate in one-way QC using the quantum circuit representation and the cluster state representation, respectively. Before the CZ gate, the $q$ and $p$ quadratures for the mode 1(2) are described as $(\hat{q}_{1(2)}, \hat{p}_{1(2)})$, respectively. The CZ gate transforms the quadratures as $(\hat{q}_{1(2)}, \hat{p}_{1(2)}+\hat{q}_{2(1)})$.
When the input mode 1 is measured in the $p$ quadrature, the quadratures for mode 2 are transformed to $(\hat{p}_1, \hat{p}_2+\hat{q}_1)$ after the feed-forward operation in the $q$ quadrature for the mode 2.
In the case of infinite squeezing, i.e., $\langle(\Delta \hat{p}_2)^2\rangle \to 0$, the quadratures $(-\hat{p}_1, \hat{p}_2+\hat{q}_1)$ become $(-\hat{p}_1, \hat{q}_1)$.
Thus, the above circuit with the infinite squeezing corresponds to the Fourier gate $\hat{R}(\pi/2)$ described in Eq.~(\ref{rotation}).
In a practical implementation, due to a finite squeezing effect in a practical implementation, the state after this circuit suffers from a noise corresponding to $\langle(\Delta \hat{p}_2)^2\rangle=\frac{1}{2}e^{-2r}$.
For the implementation of the other quantum gates, it is known that an arbitrary single-mode Gaussian gate can be achieved by using a four-mode linear cluster state via changing the measurement basis~\cite{ukai2010universal}.
For example, given the measurement on the $i$-th mode in the quadrature $\hat{p}+m_i\hat{q}$, the phase gate is implemented by four measurements with $(m_1,m_2,m_3,m_4)=(1,0,0,0)$~\cite{ukai2010universal, menicucci2014fault}.
In Sec.~\ref{subsubsec:building_blocks}, the experimental realization of the single- and two-mode gates in one-way CVQC is described.
To achieve universality, the cubic phase gate should be implemented within the framework of one-way QC.
Figure~\ref{mbqcgate}(c) shows the quantum circuit representation for the cubic phase gate based on the measurement-induced scheme using the cubic phase state $\ket{\gamma}$.
Figure~\ref{mbqcgate}(d) shows the cluster state representation of Fig.~\ref{mbqcgate}(c)~\cite{06Menicucci,09Gu}.

\begin{figure}[t]
 \centering \includegraphics[angle=0, scale=1.0]{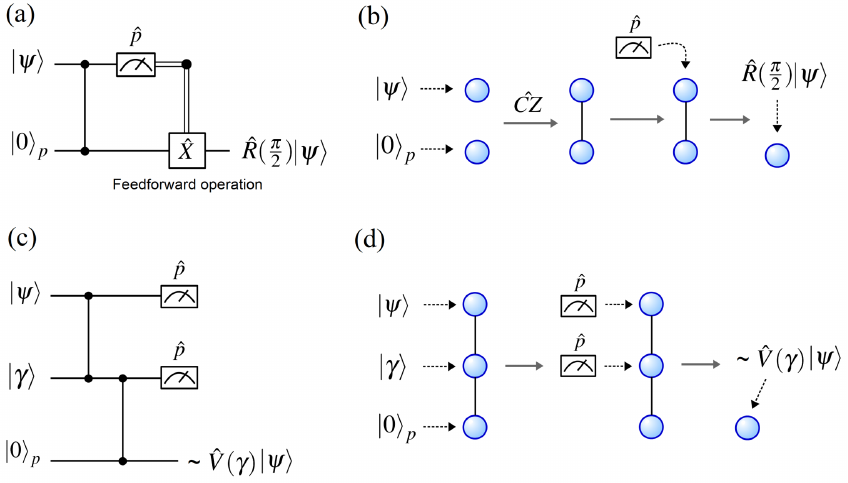} 
\caption{Quantum gates in one-way QC. (a) Quantum circuit for the Fourier gate. (b) Cluster state representation for the Fourier gate. (c) Quantum circuit for the cubic phase gate. (d) Cluster state representation for the cubic phase gate~\cite{09Gu}.}
\label{mbqcgate}
\end{figure}
\section{Scalable quantum computation with CVs}

\subsection{Scalability of photonic QC}

\subsubsection{Building blocks for photonic QC}\label{subsubsec:building_blocks}

Recent experimental progress has found unique strategies that make optical CVQC highly scalable.
Before discussing such strategies, let us first briefly review the history of developing
basic building blocks for optical CVQC over the past several decades~\cite{09OBrien,15Andersen,19Takeda2,19Pfister}.

One of the most essential building blocks for optical QC is quantum light sources.
The most commonly used quantum light source in CV systems is 
the sources for squeezed vacuum states $\hat{S}(r)\ket{0}$
because these states not only well approximate the quadrature eigenstate $\ket{0}_q$ in the high squeezing limit ($r\to\infty$)
but also play an important role in generating various CV entangled states~\cite{16Andersen}. 
The first generation of squeezed light in 1985 used four-wave mixing in an atomic vapor of sodium atoms~\cite{85Slusher}.
However, it was later found that high-level squeezed light can be generated more conveniently by an optical parametric oscillator (OPO), which is an optical cavity with a second-order nonlinear crystal inside and pumped by a continuous-wave beam~\cite{86Wu}.
In this configuration, more than 10 dB (up to 15dB) of squeezing has been observed~\cite{vahlbruch2008observation,eberle2010quantum,16Vahlbruch}.
Thus far we have only focused on single-mode squeezing produced from OPOs operated below threshold. However, depending on the configuration and operating condition, OPOs can also produce multimode squeezing and entanglement, such as standard two-mode squeezing~\cite{ou1992realization},
entanglement between spatial or temporal modes~\cite{lassen2009continuous,chalopin2010multimode,zambrini2003polarization,gatti1997langevin},
frequency modes~\cite{14Chen,14Roslund,17Cai},
and different colors~\cite{villar2005generation,coelho2009three}.
The OPOs are also used to generate non-Gaussian states with high purity,
such as single photon states~\cite{07Nielsen}, Fock-state superposition states~\cite{13Yukawa}, and Schr\"{o}dinger's cat states~\cite{06Nielsen,07Wakui, 08Takahashi},
by sending part of the squeezed light to photon detectors and adopting specific detection events.

\begin{figure}[!t]
\begin{center}
\includegraphics[width=\linewidth,clip]{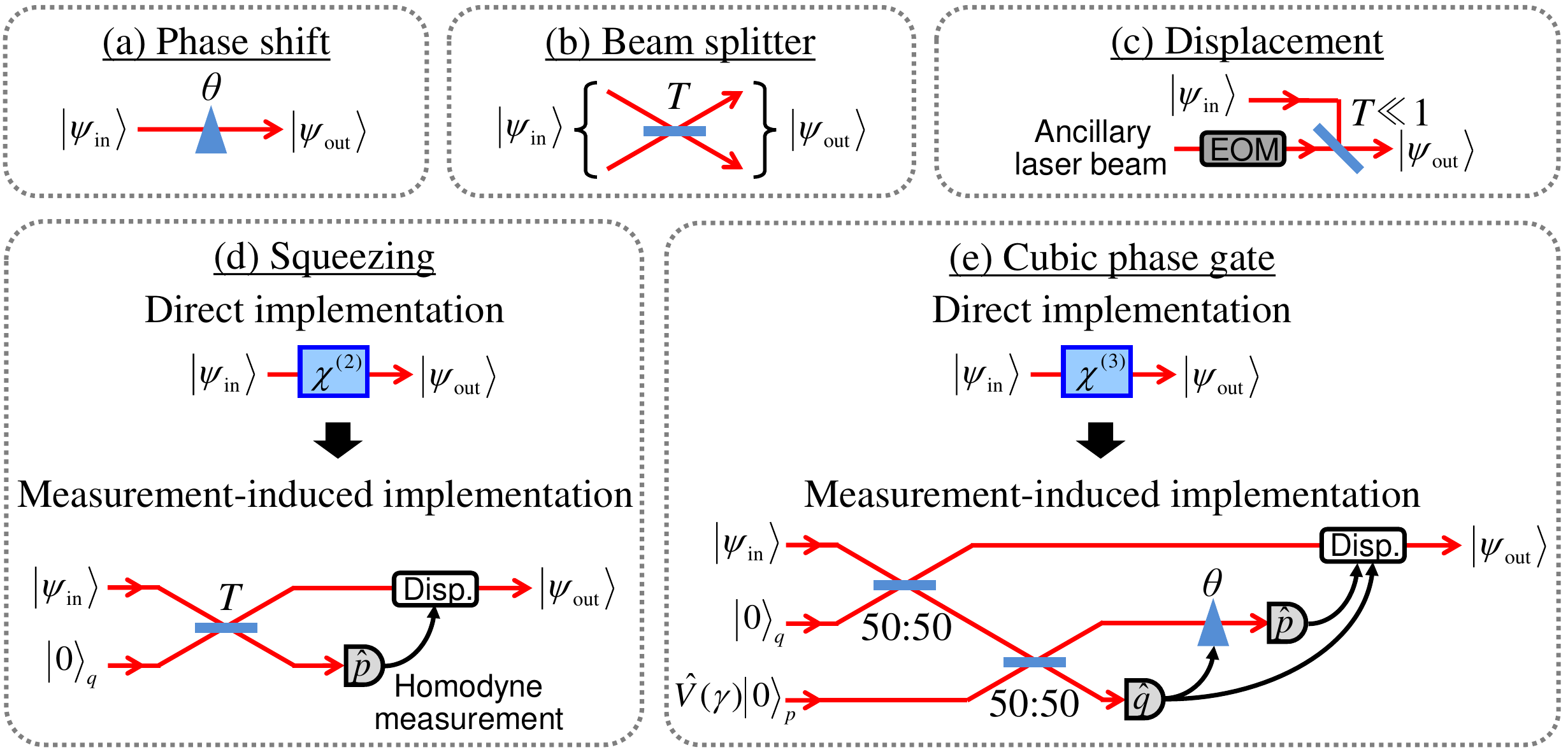}
\end{center}
\caption{Optical implementation of a universal CV gate set.
$\chi^{(2)}$ and $\chi^{(3)}$ represent second- and third-order nonlinearity, respectively.
Measurement-induced implementations are based on Refs.~\cite{05Filip,16Miyata}.
EOM, electro-optic modulator; Disp., displacement operation.
}
\label{fig:GateImplementation}
\end{figure}
The central challenge in optical QC has been the realization of a universal set of CV quantum gates,
the combination of which allows us to perform an arbitrary unitary transformation on optical quantum states (Fig.~\ref{fig:GateImplementation}).
From the experimental point of view,
here we consider a universal set composed of four Gaussian gates,
including phase shift $\hat{R}(\theta)$,
beam splitter $\hat{BS}(\theta)$,
displacement $\hat{D}(\alpha)$,
and squeezing operations $\hat{S}(r)$,
as well as at least one non-Gaussian gate such as a cubic phase gate $\hat{V}(\gamma)$
(their definitions are given in Sec.~\ref{subsec:cvqc}).
Deterministic implementation of all these Gaussian gates has already been well-established.
Phase shift and beam splitter operations are easy to implement with only passive linear optics (Figs.~\ref{fig:GateImplementation}(a) and (b)).
Displacement operations are also easily implemented by partly mixing a modulated ancillary beam to the target beam (Fig.~\ref{fig:GateImplementation}(c)).
In contrast, implementing squeezing operations requires second-order nonlinearity and thus is nontrivial.
A straightforward way to perform squeezing is to directly couple the target beam to a nonlinear medium,
but this method often suffers from unwanted coupling loss that degrades the operational fidelity.
This problem is avoided by a measurement-induced squeezing gate~\cite{05Filip},
where an ancillary squeezed vacuum state is consumed to indirectly apply a squeezing operation to the target state (Fig.~\ref{fig:GateImplementation}(d)).
Such a measurement-induced squeezing gate has been demonstrated~\cite{07Yoshikawa}
and further extended to implement a two-mode CX gate for CVs~\cite{08Yoshikawa}.
In the early experiments, these gates were performed on Gaussian input states defined in optical narrow frequency sidebands. They were later updated to cover much broader bandwidth so that they can be applied to non-Gaussian input states that are usually created in optical wave packets with broad frequency spectra~\cite{14Miwa, 18Shiozawa}.

Instead of constructing these quantum gates one by one in optical circuits,
one-way QC based on CV cluster states has also been pursued as an alternative approach.
In the early experiments, small-scale CV cluster states with up-to eight modes have been generated by preparing multiple squeezed light sources in parallel and mixing the generated squeezed vacua with multiple beam splitters~\cite{07Su, 08Yukawa,12Su}.
Gaussian quantum gates, including single-mode~\cite{11Ukai}, two-mode~\cite{11Ukai2}, and also sequential gates~\cite{13Su}, were demonstrated by performing homodyne measurements on these cluster states with appropriate measurement bases.

Thus far, all demonstrated CV gates were limited to Gaussian gates.
Non-Gaussian gates have yet to be achieved, and they have been one of the most challenging building blocks in optical CVQC.
However, there has been continuous theoretical and experimental progress towards the realization of non-Gaussian gates based on measurement-induced schemes.
As already mentioned in Sec.~\ref{subsubsec:onewaygates},
Ref.~\cite{01Gottesman} proposed an original measurement-induced scheme for deterministic cubic phase gates $\hat{V}(\gamma)$,
requiring only Gaussian operations except for the preparation for an ancillary cubic phase state
$\hat{V}(\gamma)\ket{0}_p$.
The original proposal was later rearranged to a simpler implementation method which is experimentally more feasible~\cite{11Marek,16Miyata} (Fig.~\ref{fig:GateImplementation}(e)).
Moreover, it was also found that higher-order phase gates can be implemented similarly with additional non-Gaussian ancillary states~\cite{18Marek}.
A probabilistic source of approximated cubic phase states~\cite{13Yukawa}
and the feedforward system for the cubic phase gate~\cite{14Miyata} have been reported already.
Toward deterministic cubic phase gates, an all-optical quantum memory to
store the cubic phase states has also been investigated~\cite{19Hashimoto}.
These technologies are expected to lead to the demonstration of a cubic phase gate shortly.

\subsubsection{Integrating building blocks to scale up QC}

Once all the necessary basic building blocks for photonic QC are realized,
it is in principle possible to scale up photonic QC by building large optical circuits incorporating all these ingredients.
However, this kind of in-principle scalability does not mean that the implementation is experimentally efficient.
All the early proof-of-principle demonstrations introduced in Sec.~\ref{subsubsec:building_blocks}
used bulky optical components in free space and beam-path encoding where one beam path represents one quantum mode.
Such implementations are flexible and suitable for building simple optical circuits designed for specific purposes.
However, the number of optical components and spaces linearly increases with the number of optical modes and operational steps.
A vast number of gates and modes are estimated to be required
for quantum computers to outperform current supercomputers in solving practical and meaningful quantum algorithms~\cite{jones2012layered,21Gidney}.
In Ref.~\cite{21Gidney}, the number of the physical gates and modes for factoring 2048 bit integers are estimated as $\sim10^{9}$ and $2\times10^{7}$, respectively.
Building such a powerful quantum computer by the straightforward extension of the previous implementations is almost impractical.

However, recent progress has revealed that the property of light enables us to overcome the problem and efficiently scale up photonic QC.
Here, we focus on three important research directions to scale up photonic QC:
one is optical multiplexing techniques to use the rich degrees of freedom of light;
second is the increase in operational bandwidth to utilize the large information capacity of light;
third is photonic chips to miniaturize and integrate optical components.
In the following Secs.~\ref{subsec:multiplexing} to \ref{subsec:Integrated},
we introduce the basic idea and recent experimental progress for each direction.

\subsection{Scaling up with multiplexing}\label{subsec:multiplexing}

\subsubsection{Types of multiplexing}

The basic implementation of optical quantum information processing uses path encoding. However, there exist many other degrees of freedom of light for encoding, such as frequency, time, and spatial modes. These degrees of freedom enable us to potentially pack a large amount of quantum information in a single optical path and perform large-scale QC more efficiently. This approach is called multiplexing and one of the unique advantages in optical systems. We introduce three different options for multiplexing below, each of which has advantages and disadvantages. Note that these options can be combined and simultaneously exploited.

The first option is spatial-mode multiplexing using different spatial distributions of light.
In principle, one can define arbitrarily many orthogonal spatial modes in a single optical beam.
For example, the experiment in Ref.~\cite{12Armstrong}
generated multimode CV entangled states in a single optical beam by
defining modes to be combinations of different spatial regions of one beam.
In another approach,
several spatial modes with different orbital angular momentum (OAM) were used
to generate entangled states~\cite{09Lassen,20Wang}
or realize quantum teleportation of several modes in parallel~\cite{20Liu}.
These approaches may be suitable for
enhancing the data-carrying capacity in quantum communication protocols
by combining multiple channels into a single optical transmission channel.
However, it is difficult to individually access or separate each spatial mode in a single beam, and thus the efficient scheme to implement QC in this strategy is unclear.

The second option is frequency multiplexing using different frequency bins.
One prominent approach in this direction is to use
equally spaced resonant frequency modes (optical frequency combs) of an OPO~\cite{04Olivier,08Menicucci,19Pfister}.
By pumping a single OPO with multi-frequency continuous-wave pump beams,
up to 60 frequency modes were demonstrated to be entangled~\cite{14Chen}.
In contrast, another approach repeatedly pumps an OPO by synchronized pulsed pump beams,
succeeding in generating entangled states of $\sim10$ independent frequency modes~\cite{14Roslund,17Cai}
and non-Gaussian states in selectable frequency modes~\cite{20Ra}.
The challenges in these approaches include the limitation of maximum frequency modes
(for example, the phase-matching bandwidth limits the modes in the frequency-comb approach on the order of $10^4$~\cite{19Pfister})
as well as difficulties in frequency-sensitive measurement
and efficient separation of individual frequency modes.

The third option is time multiplexing using different time bins~\cite{19Takeda2}.
By dividing an optical beam into non-overlapping time bins,
an unlimited number of wave-packet modes can be defined.
These modes are individually accessible, and the same optical components can be repeatedly used at different times for operations and measurements of the modes.
Time multiplexing is currently one of the leading approaches in optical CVQC,
as highlighted by recent demonstrations of ultra-large-scale cluster states and one-way QC with these states~\cite{13Yokoyama, 16Yoshikawa, 19Asavanant, 19Larsen, 20Asavanant, 20Larsen}.
One disadvantage of this approach is the lossy long optical delay lines required for introducing interaction between optical pulses at different times and increasing the number of processable modes.
In the following Secs.~\ref{subsubsec:time-one-way} and \ref{subsubsec:loop}, we focus on 
two specific approaches in this direction and explain more details
of technical progress and challenges.

\subsubsection{Time-multiplexed one-way QC}\label{subsubsec:time-one-way}

\begin{figure*}[!t]
\begin{center}
\includegraphics[width=0.8\linewidth,clip]{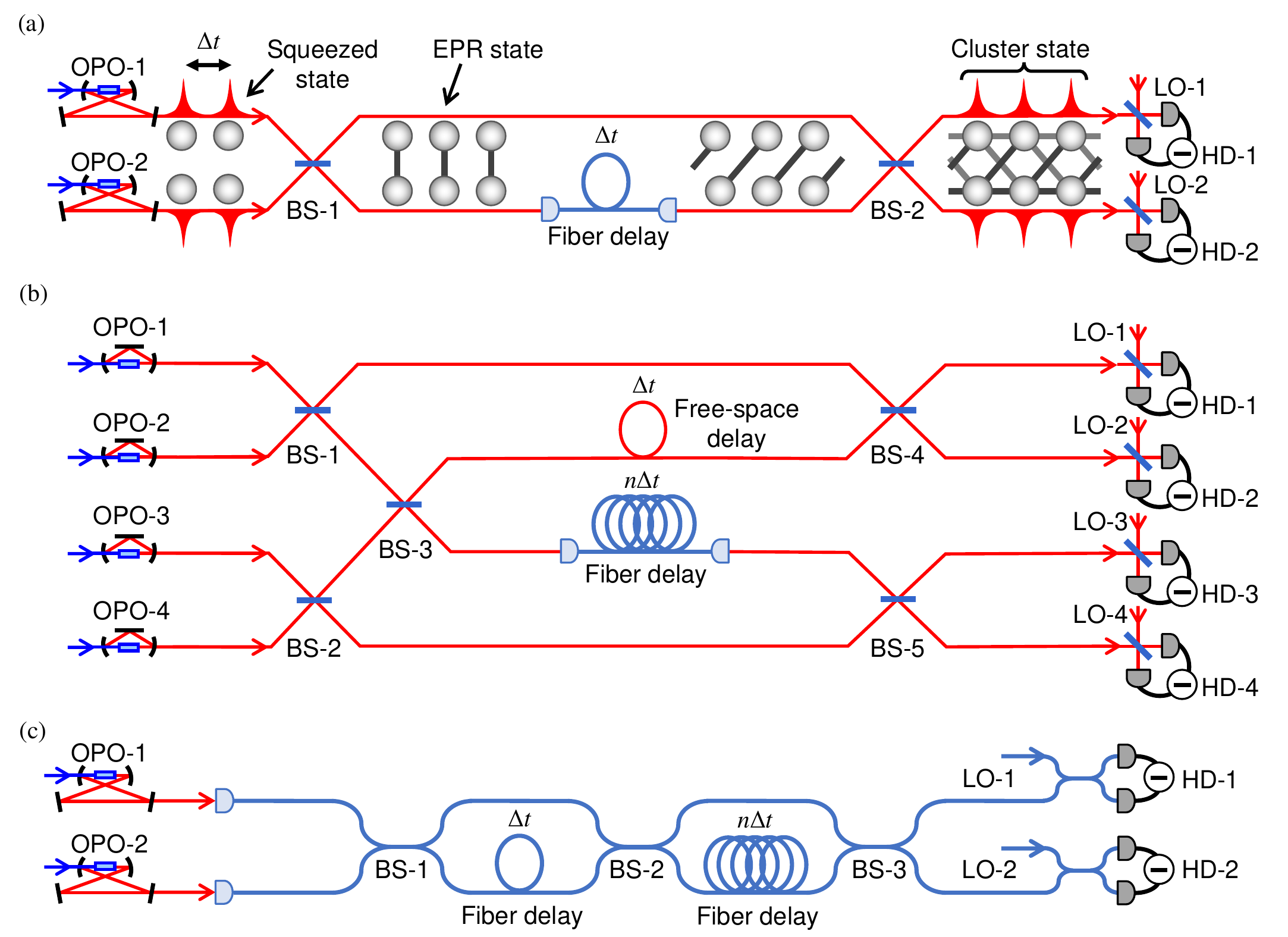}
\end{center}
\caption{Experimental schematic for generating time-multiplexed cluster states.
(a) One-dimensional cluster states in Refs.~\cite{13Yokoyama,16Yoshikawa}.
(b) Two-dimensional cluster states in Ref.~\cite{19Asavanant}.
(c) Two-dimensional cluster states in Ref.~\cite{19Larsen}.
OPO, optical parametric oscillator; BS, beam splitter; HD, homodyne detector; LO, local oscillator.
}
\label{fig:ClusterGeneration}
\end{figure*}

Here, we review the recent impressive experimental progress in one-way CVQC with a time-multiplexing approach.
As mentioned in Sec.~\ref{subsubsec:building_blocks},
the early demonstrations of one-way QC relied only on path encoding~\cite{07Su, 08Yukawa,12Su}.
However, this encoding requires one squeezed light source for one quantum mode, making it infeasible to generate
large-scale cluster states.
In around 2010, novel ideas to efficiently generate large-scale cluster states with time multiplexing were proposed~\cite{10Menicucci, 11Menicucci}.
Based on these ideas, the generation of more than 10,000-mode cluster states was verified in 2013~\cite{13Yokoyama}.
Later the number of modes was increased up to one million by technical improvement of the experimental system~\cite{16Yoshikawa}.
In fact, this is the largest number of modes confirmed to be entangled in any physical system.
The schematic of these experiments is shown in Fig.~\ref{fig:ClusterGeneration}(a).
In this setup, two OPOs continuously generated squeezed beams,
each of which was divided into time bins to define a train of pulsed squeezed states.
These squeezed states were then combined at the first beam splitter to produce
an Einstein-Podolsky-Rosen (EPR) entangled state.
The optical delay line followed by the second beam splitter combined
one part of the EPR state with one part of another EPR state.
This interaction finally generated a large-scale entangled state,
where quantum modes were connected in a one-dimensional chain fashion.
The generated states were proven to be equivalent to one-dimensional CV cluster states,
a resource for one-input one-output QC.

\begin{figure}[!t]
\begin{center}
\includegraphics[width=\linewidth,clip]{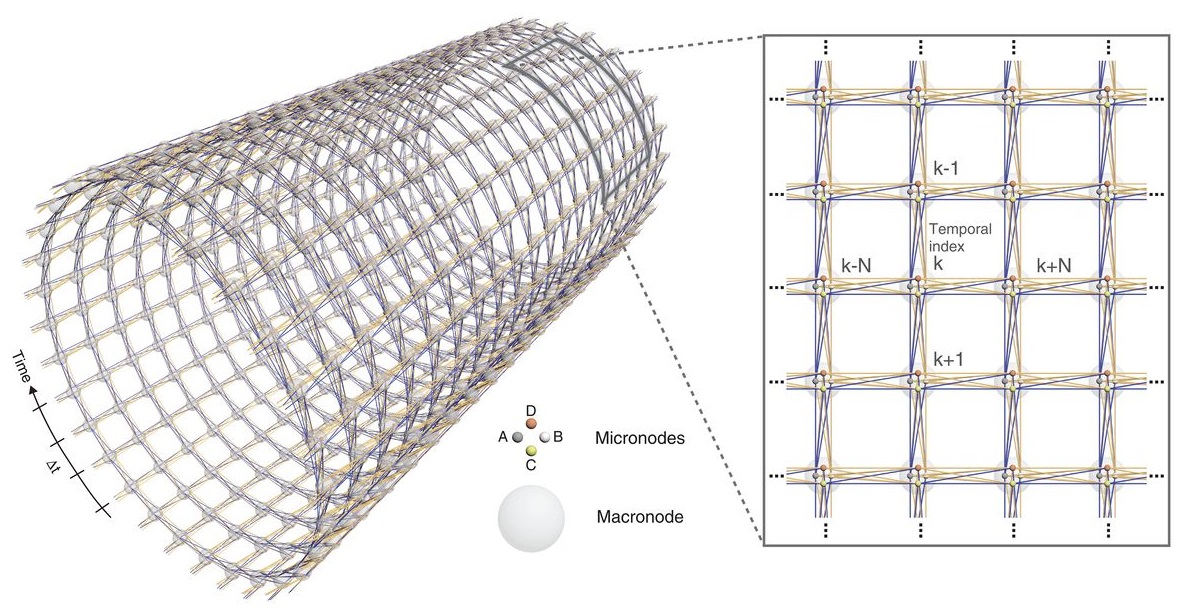}
\end{center}
\caption{Graphical representation of the two-dimensional CV cluster states.
Reproduced from \cite{19Asavanant}.
}
\label{fig:Cylinder}
\end{figure}

In 2019, two research groups further extended this approach to generate two-dimensional cluster states, a universal resource for multi-input multi-output one-way QC~\cite{19Asavanant, 19Larsen}.
Although experimental setups were slightly different (Figs.~\ref{fig:ClusterGeneration}(b) and (c)), these two demonstrations share the common idea that two optical delay lines with different lengths allow for entangling optical modes in two dimensions.
The shorter delay line was for connecting neighboring modes to produce one-dimensional cluster states
and the longer one was for connecting these cluster states in another dimension to produce two-dimensional cluster states.
The graph of the resultant cluster state can be depicted as quantum modes arranged on a continuous cylindrical structure in both experiments, as shown in Fig.~\ref{fig:Cylinder}.
The number of modes along the cylinder circumference was determined by the length of the longer delay line and limits the number of input modes for QC. On the other hand, the number of modes along the cylindrical axis was infinite in principle, enabling an unlimited number of quantum operations.

Recently, these groups have demonstrated one-way QC using these time-multiplexed cluster states. CV one-way QC can be performed by homodyne measurement of each mode of cluster states with appropriate measurement basis and the following displacement operation on specific modes.
When using time-multiplexed cluster states, it is necessary to change the measurement basis pulse-by-pulse in time.
In fact, rapidly changing the homodyne measurement basis is easily achieved by phase-modulating the LO beams with electro-optic modulators. The phase modulation pattern can be defined by programmable electric signals, meaning that the gate sequence can be easily programmed and changed according to computational purposes. It is an advantage over the standard circuit-model QC whose gate sequence is determined by the arrangement of optical components and thus not easily variable. Thus far, one-input 100-step quantum gates with one-dimensional cluster states~\cite{20Asavanant} and three-input 12-step quantum gates with two-dimensional cluster states~\cite{20Larsen} have been experimentally demonstrated. In these experiments, optical displacement operations were replaced by classical signal processing after the measurement.
This replacement is valid when the purpose is to know only classical calculation results at the final step.

Despite these impressive experiments demonstrating scalability,
there still exist a lot of challenges to be overcome in time-multiplexed one-way QC.
First, the current implementations rely on optical delay lines to increase the number of processable modes.
An optimal choice for long delay lines is optical fibers having the minimum loss of 0.2 dB/km at telecommunication wavelength.
For example, transmitting a 1-km optical fiber introduces 5\% loss, which is below the fault-tolerant threshold for some QEC schemes~\cite{18Fukui}.
The minimum time interval of wave packets in the recent experiments is 40 ns (12 m in length)~\cite{20Asavanant},
and thus the 1-km fiber enables QC with the order of 100 processable modes.
Further scale-up requires a shorter time interval or modified architectures
(one methodology to shorten the time interval is given in Sec.~\ref{subsec:broadband}).
Another issue is that all the previous implementations have been limited to the Gaussian regime, which is efficiently simulatable by classical computers~\cite{02Bartlett}.
Possible options for introducing non-Gaussian elements include photon-number resolving measurement on cluster states~\cite{06Menicucci,09Gu}
or injection of non-Gaussian states as ancillary states~\cite{18Alexander}.
Finally, QEC needs to be incorporated into the one-way QC.
For this purpose, the extension from two- to three-dimensional cluster states
and the introduction of bosonic QECCs are being considered~\cite{20Fukui,20Wu,21Bourassa,21Larsen}.

\subsubsection{Loop-based architecture for time-multiplexed QC}\label{subsubsec:loop}

Another emergent approach for scalable time-multiplexed optical QC
is to use loop-based architectures, where optical pulses circulating in loops transmit the same optical components repeatedly.
The repeated use of the same optical components leads to a dramatic reduction of the circuit size, making photonic QC more scalable.
In 2014, Ref.~\cite{14Motes} proposed a dual-loop architecture 
that realizes an arbitrary multi-mode linear interferometer in a compact setup.
This architecture was proven to have applications in
boson sampling and universal QC based on single photons~\cite{14Motes,15Rohde},
and several related experiments have been reported already~\cite{55Schreiber,17He}.

\begin{figure}[!t]
\begin{center}
\includegraphics[width=\linewidth,clip]{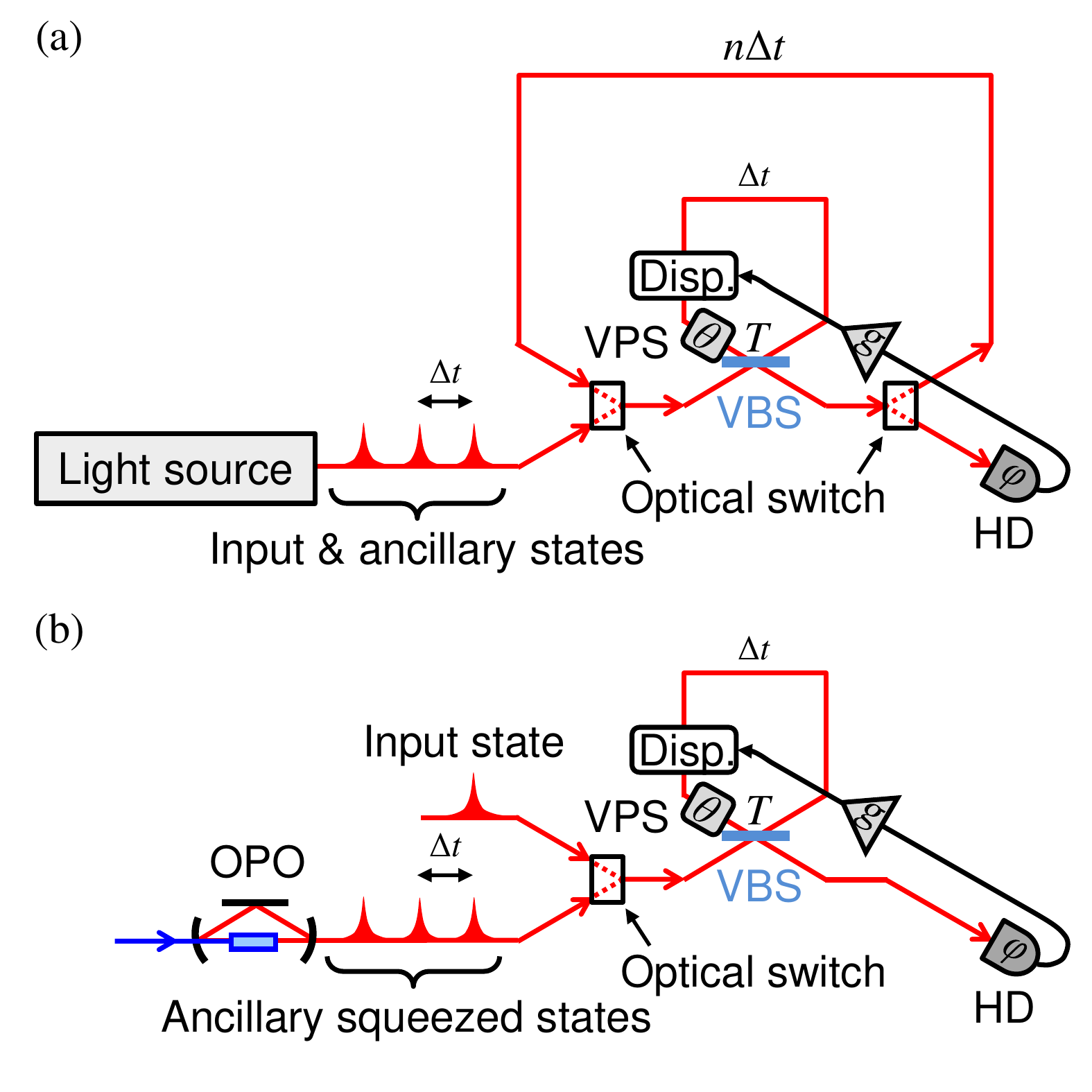}
\end{center}
\caption{Loop-based architecture for CVQC.
(a) A universal multi-mode quantum processor proposed in Ref.~\cite{17Takeda}.
(b) A single-mode quantum processor demonstrated in Ref.~\cite{21Enomoto}.
EOM, electro-optic modulator; VPS, variable phase shifter; VBS, variable beam splitter.
}
\label{fig:LoopCircuit}
\end{figure}

This idea was later extended to a dual-loop architecture for universal CVQC in Fig. ~\ref{fig:LoopCircuit}(a)~\cite{17Takeda}.
The essential working principle of this architecture is simple.
The outer loop is a quantum memory to store a train of optical pulses for input states and ancillary states for QC. The inner one is a quantum processor to sequentially perform gates to the pulses by dynamically controlling several parameters, such as beam splitter transmissivity~$(T)$, switch polarity, phase shift~$(\theta)$, measurement basis~$(\phi)$, and feedforward gain~$(g)$. This configuration enables us to perform a universal set of gates either directly or by measurement-induced schemes~\cite{05Filip,16Miyata}, once necessary ancillary states are provided. This architecture can deal with arbitrarily many modes and operational steps without increasing the number of optical components. Moreover, the gate sequence can be easily modified by changing the control signal of the parameters, meaning that the QC is programmable.

One advantage of the above loop-based scheme compared to the one-way QC scheme in Sec~\ref{subsubsec:time-one-way} is the decrease in the required resources.
One-way QC first prepares universal entanglement resources for arbitrary QC.
When performing a specific computational task, unnecessary parts of the entanglement have to be removed by measurement and feedforward operations~\cite{18Alexander,10Miwa}.
This process increases operational steps and often introduces additional noise and imperfection to the computation.
In contrast, the loop-based scheme requires fewer steps because it creates entanglement only when necessary.
However, the challenge is the requirement for fast and precise control of many variable parameters,
which is unnecessary for the one-way QC.

Recent efforts have tackled the technical challenge and partly demonstrated the advantageous functionalities of the dual-loop architecture.
Based on the original proposal,  Ref.~\cite{19Takeda} dynamically controlled a single-loop circuit with a variable beam splitter and phase shifter to demonstrate the programmable and scalable generation
of various entangled states from two-mode EPR states to 1,000-mode cluster states.
A later experiment updated this circuit to a single-mode photonic quantum processor in Fig. ~\ref{fig:LoopCircuit}(b),
demonstrating programmable and multi-step Gaussian operations~\cite{21Enomoto}.
This demonstration included optical feedforward that was omitted in
the time-multiplexed one-way QC experiments~\cite{19Asavanant, 19Larsen}.
This inclusion of feedforward means that this processor can finalize quantum operations and export the output optical quantum states for further use, thereby working as a versatile photonic quantum processor with potential applications to quantum communication and sensing.
The next challenge will be the demonstration of non-Gaussian gates and multi-mode gates.

The idea of using loop structures has been introduced in several proposals for optical CV quantum information processing.
In contrast to the dual-loop architecture in Ref.~\cite{17Takeda}, the chain-loop architecture was introduced in Ref.~\cite{18Qi} to
implement universal QC with lower loss.
The loop structure is also effective for increasing the success probability of photon subtraction~\cite{18Marek} and efficient implementation and verification of Gaussian boson sampling~\cite{20Abrahao}.

\subsection{Scaling up with broadening bandwidth}\label{subsec:broadband}

\subsubsection{Why broadband?}

The carrier frequency of a light field is a few hundred THz, and 
the available broad frequency space is a rich resource
to encode large-capacity quantum information.
This resource will ultimately enable us to process
unlimitedly many independent frequency modes simultaneously in frequency multiplexing
or ultra-short optical pulse trains at an ultra-high clock frequency in time multiplexing.
However, the operational bandwidth in actual optical QC is inherently limited to
the electronically accessible MHz-to-GHz range, posing a dramatic gap between
potentially usable bandwidth and accessible bandwidth.
For example, the maximum bandwidth in recent experiments on CV quantum operations
is $\sim100$ MHz~\cite{18Shiozawa},
limiting the operational clock frequency in time-multiplexed processing to 25 MHz at most~\cite{20Asavanant}.
This limitation came from the bandwidth of squeezed light sources
and electronics to measure and manipulate optical states. 
Broadening the bandwidth of these elements is an important research direction
to exploit the optical bandwidth maximally.
Below, we refer to recent advancements in this direction
and finally explain, as an ultimate dream, an all-optical QC scheme that is not limited by electronic bandwidth anymore.

\subsubsection{Ultra-broadband squeezed light source}

One of the most fundamental light sources in CVs
is squeezed light sources, which are used for producing various Gaussian and non-Gaussian states.
The bandwidth of the squeezed light sources usually limits the bandwidth of the generated optical states and their generation rates. 
As already mentioned in Sec.~\ref{subsubsec:building_blocks},
the most successful squeezed light source to date is an OPO,
where second-order nonlinearity is enhanced in optical cavities.
Instead, these cavities limit the bandwidth of the generated squeezed states.
For example, typical ring-cavity OPOs used in state-of-the-art time-multiplexed experiments~\cite{19Asavanant,20Asavanant,19Takeda,21Enomoto}
have only 65 MHz bandwidth~\cite{16Serikawa}, which mainly limited the computational speed of these experiments.
The bandwidth can be broader by fabricating smaller cavities but has been limited to around 2 GHz even with a monolithically integrated OPO~\cite{13Ast}.

\begin{figure}[!t]
\begin{center}
\includegraphics[width=\linewidth,clip]{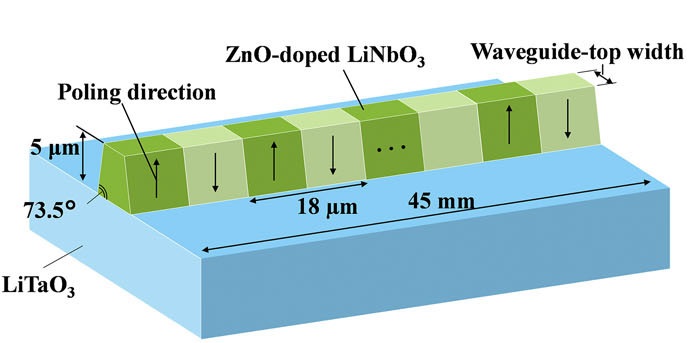}
\end{center}
\caption{Design of the periodically poled LiNbO$_3$ waveguide for OPA.
Reproduced from \cite{20Kashiwazaki}.
}
\label{fig:WaveguideOPA}
\end{figure}

This bandwidth limit can be broken by removing cavities and using a single-pass optical parametric amplifier (OPA).
The OPA's bandwidth is no longer limited by the cavities and reaches THz, limited only by dispersion or phase-matching conditions of the nonlinear crystal itself.
In general, high-level squeezing cannot be produced by
simply injecting a continuous-wave pump beam into a standard nonlinear crystal without cavity enhancement.
Instead, pulsed pump beams with intense peak power can induce sufficient nonlinearity in this single-pass OPA~\cite{87Slusher}.
However, the pulsed configuration introduces another technical difficulty in optimizing spatiotemporal mode matching between pulsed beams and thus often degrades the quality of the squeezed light~\cite{94Kim}.
Another option is to use waveguide nonlinear crystals, enhancing nonlinearity
by transverse field confinement of pump beams in small cross-sections over long interaction length.
Several experiments have reported squeezed light generation from waveguide OPAs with continuous-wave pump beams~\cite{07Yoshino,16Kaiser},
but until recently, the squeezing level was limited to around 2 dB~\cite{19Mondain,09Pysher}.
In 2020, 4.0-dB squeezing was reported by a fiber-coupled OPA module with a periodically poled LiNbO$_3$ (PPLN) waveguide~\cite{20Takanashi}.
Without the fiber coupling, the same waveguide produced 6.3-dB squeezing and
was confirmed to have the squeezing bandwidth of 2.5 THz~\cite{20Kashiwazaki} (Fig.~\ref{fig:WaveguideOPA}).
In these experiments, a higher level of squeezing is expected by improving the fabrication process of the waveguides
to reduce the optical propagation loss due to structural imperfections.
Such broadband squeezed light sources will further strengthen the capacity of frequency- and time-multiplexing.
For example, their THz-order bandwidth enables us to use micrometer wave-packet modes for time multiplexing, much shorter than the tens-of-meter modes in recent experiments~\cite{19Asavanant,19Larsen,20Asavanant,20Larsen,19Takeda}. 
It dramatically reduces the length of required delay lines and downsizes the optical circuits, ultimately leading to large-scale time-multiplexed QC on compact photonic chips.

\subsubsection{Ultra-broadband quadrature measurement}

\begin{figure*}[!t]
\begin{center}
\includegraphics[width=\linewidth,clip]{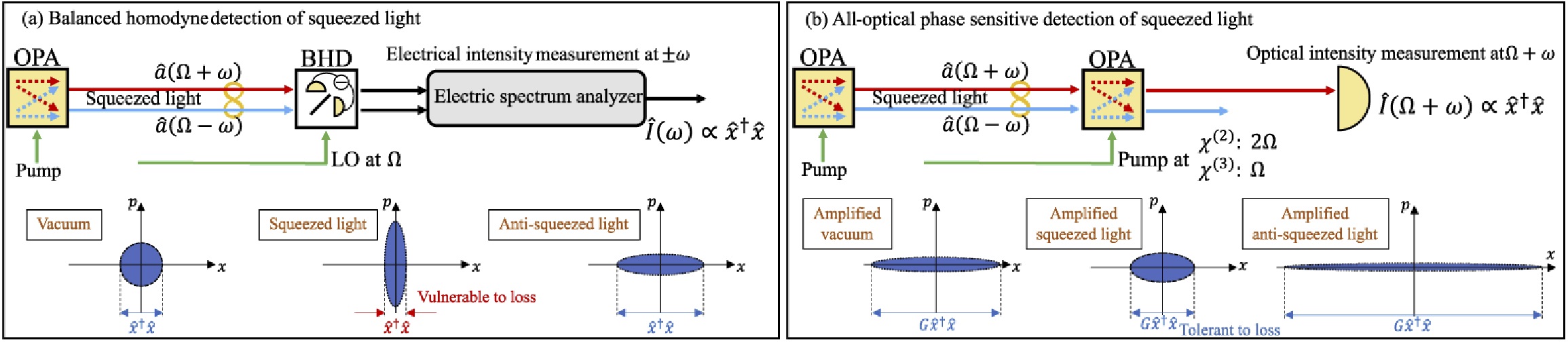}
\end{center}
\caption{Standard homodyne detection (a) and
OPA-based quadrature measurement (b).
Reproduced from \cite{20Takanashi2}.
}
\label{fig:OpticalHomodyne}
\end{figure*}

In addition to the broadband light sources, broadband light detectors are required
for broadening the operational bandwidth of photonic QC.
In CVs, the most commonly used detector
is homodyne detectors to measure quadratures of light fields.
Among the electronics in CVQC,
the bandwidth of the homodyne detector most severely limits the overall operational bandwidth
(other signal processing, such as amplification and modulation, can easily reach GHz-bandwidth operation).
Standard homodyne detection lets an input optical signal
interfere with a LO beam,
converts the optical signal into photocurrent by photodiodes,
and then amplifies the signal by transimpedance amplifiers.
The response of photodiodes and amplifiers usually limit the speed performance of the detector.
Conventional high-quantum-efficiency homodyne detectors for quantum optics have
typically had bandwidths of $\sim100$ MHz or below.~\cite{11Chi,12Kumar}.
Recently, high-speed homodyne detectors with up to 1.1-GHz bandwidth have been reported
with free-space photodiodes~\cite{18Zhang}.
Another experiment has developed a 1.7-GHz bandwidth homodyne detector
with on-chip photodiodes and measured squeezed light with up to 9 GHz~\cite{21Tasker}.
These GHz-bandwidth homodyne detectors will lift the bandwidth of QC
from the current MHz range to the GHz range.
However, as far as electronics are concerned, a further dramatic increase in bandwidth cannot be expected.

A completely different approach to overcome this limitation is to use high-gain OPAs
for quadrature measurement, as shown in Fig.~\ref{fig:OpticalHomodyne}~\cite{18Shaked,20Takanashi2,20Jiamin}.
The parametric amplification process selects one of the quadratures of an optical field and amplifies it
without adding noise while attenuating the other quadrature.
This process is equivalent to the standard homodyne detection,
where the LO beam selects a quadrature of interest and amplifies it
to an electrical signal level sufficiently above the electric noise of the detection system.
In the high-gain OPA, a high-power pump beam plays the role of the LO beam,
amplifying the optical signal to the level sufficiently above the vacuum noise level of light.
There are several advantages to this measurement scheme.
First, the same nonlinear process can be used both for squeezed light generation and quadrature measurement,
thus guaranteeing that the measurement system can cover the entire bandwidth of the squeezed light.
Second, the high-gain OPA amplifies the originally quantum signal of quadratures
to a classical optical signal which is insensitive to optical losses, including propagation losses,
mode mismatching, and detector inefficiencies~\cite{17Mathieu,21Frascella}.
Therefore, the quadrature can be measured with high quantum efficiency
even by low-quantum-efficiency detectors.

In 2018, a high-gain OPA was realized by third-order nonlinearity of a photonic crystal fiber to achieve broadband quadrature measurement of squeezed light~\cite{18Shaked}.
In this experiment, a pulsed laser pumped the fiber to generate pulsed squeezed light through the nonlinearity. Its quadrature was then amplified by using the same nonlinear process again and finally detected by a spectrometer.
This configuration realized the observation of squeezing over 55 THz and 
also showed its robustness to detection inefficiencies.
In applications to CVQC,
a more suitable option for high-gain OPAs is to use second-order nonlinearity of
waveguide crystals with continuous-wave pump beams.
The difficulty of this approach lies in the fact that
the crystal for high-gain OPAs is required to have high optical conversion efficiency as well as
high durability to a high-power pump beam.
In 2019, an over-30-dB-gain OPA was reported by a highly durable PPLN waveguide with a continuous pump beam~\cite{19Kashiwazaki}.
In 2020, the PPLN waveguides were used both for squeezed light generation and quadrature measurement~\cite{20Takanashi2}.
As a result, 3-dB squeezing was observed over 3 THz by detecting the amplified optical signal with an optical spectrum analyzer.
The emergence of these broadband quadrature measurement schemes with high-gain OPAs
will open a promising pathway for ultrahigh capacity quantum communication and computation.

\subsubsection{Towards ultra-broadband QC by all-optical means}

As an ultimate form of ultra-broadband optical QC, there is a potential to realize ``all-optical'' QC. Its essential idea is to replace all electronics with corresponding optical components and break the electrical bandwidth limit. The resultant all-optical quantum computer may be able to exploit the entire THz bandwidth of light.

This idea dates back to the original proposal of ``all-optical'' CV quantum teleportation in 1999~\cite{99Ralph} (Fig.~\ref{fig:AllOpticalTeleportation}).
CV quantum teleportation~\cite{98Braunstein,98Furusawa} is
the fundamental building block for
one-way QC and measurement-induced quantum gates in CVs. 
The original protocol of optical CV quantum teleportation proceeds in the following four steps:
(i) generation of an EPR-entangled state shared by a sender and a receiver,
(ii) sender's joint homodyne measurement on an input state and his/her part of the entangled state,
(iii) transmission of the measured electric signals to the receiver through classical channels,
and (iv) receiver's feedforward operation to his/her part of the entangled state through electro-optic modulation.
The all-optical teleportation scheme (Fig.~\ref{fig:AllOpticalTeleportation}) replaces the homodyne detectors in step (ii) with high-gain OPAs, amplifying the optical signals without converting them to electric signals. 
The amplified optical signals can be sent through lossy classical optical channels to the receiver because they
are well above the vacuum noise level and tolerant to losses.
These optical signals are then directly used for feedforward operation
by weakly injecting the beam to the target beam via a beam splitter.
In this way, the teleportation protocol can be performed all-optically without any electronics,
leading to potentially ultra-broadband operation.

\begin{figure}[!t]
\begin{center}
\includegraphics[width=\linewidth,clip]{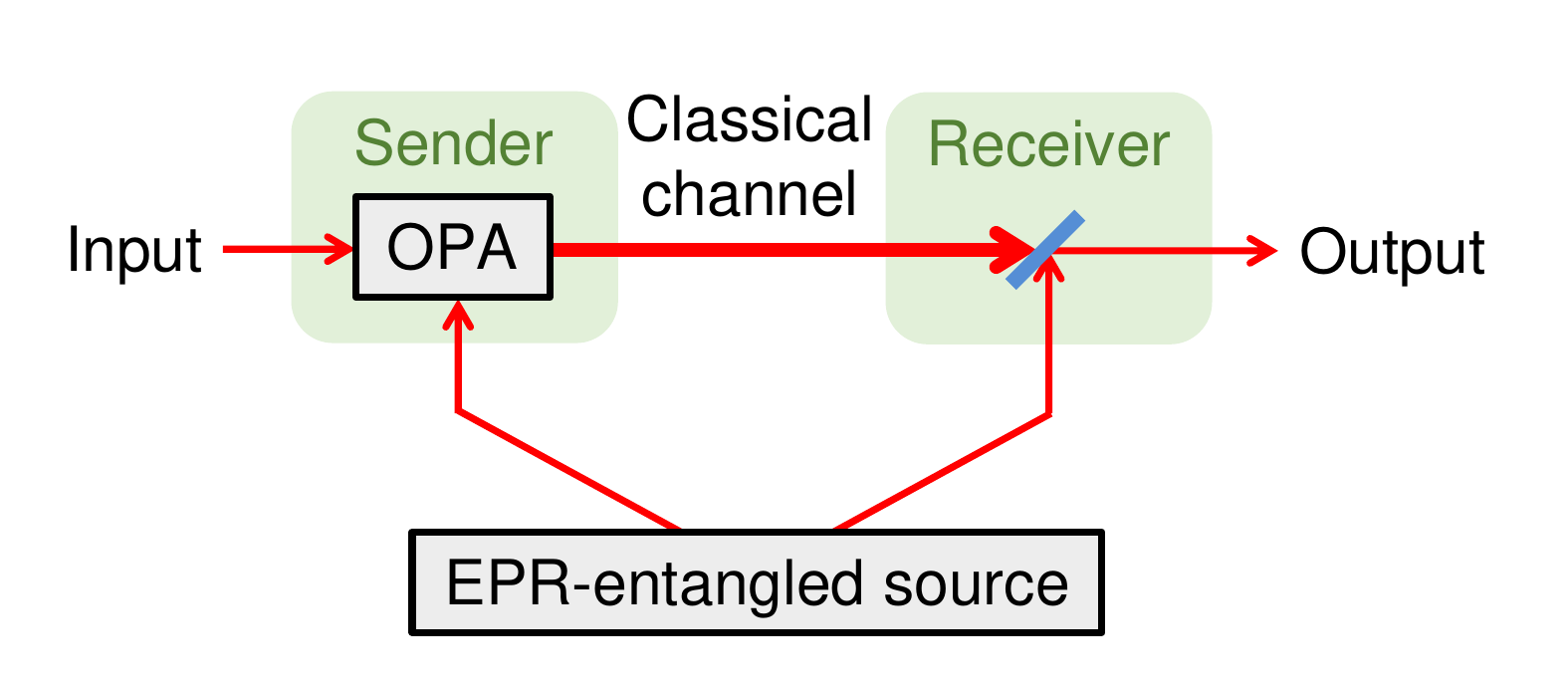}
\end{center}
\caption{All-optical quantum teleportation proposed in Ref.~\cite{99Ralph} and demonstrated in Ref.~\cite{20Liu}.
}
\label{fig:AllOpticalTeleportation}
\end{figure}

Recently, all-optical quantum teleportation was experimentally demonstrated as a means to simultaneously teleport nine channels multiplexed in the OAM of a single beam~\cite{20Liu}.
In this experiment, OPAs based on four-wave mixing were used to amplify the optical signals in different OAM modes simultaneously.
The amplified optical beam was used to perform feedforward operation directly for all nine channels in parallel.
This demonstration showed that the all-optical method is a promising path
to simultaneous processing of a large amount of quantum information encoded in degrees of freedom of a single light beam.
In terms of exploiting its frequency degree of freedom,
the recent development of ultra-broadband squeezed light sources and measurement based on waveguide OPAs~\cite{20Takanashi, 20Kashiwazaki, 20Takanashi2}
will be an enabler for ultra-broadband all-optical quantum teleportation circuits.
Such all-optical teleportation circuits can be further extended to more advanced QC protocols,
such as one-way QC and measurement-induced gate operations,
opening the path for the all-optical QC with over-THz bandwidth.

\begin{figure*}[!t]
\begin{center}
\includegraphics[width=\linewidth,clip]{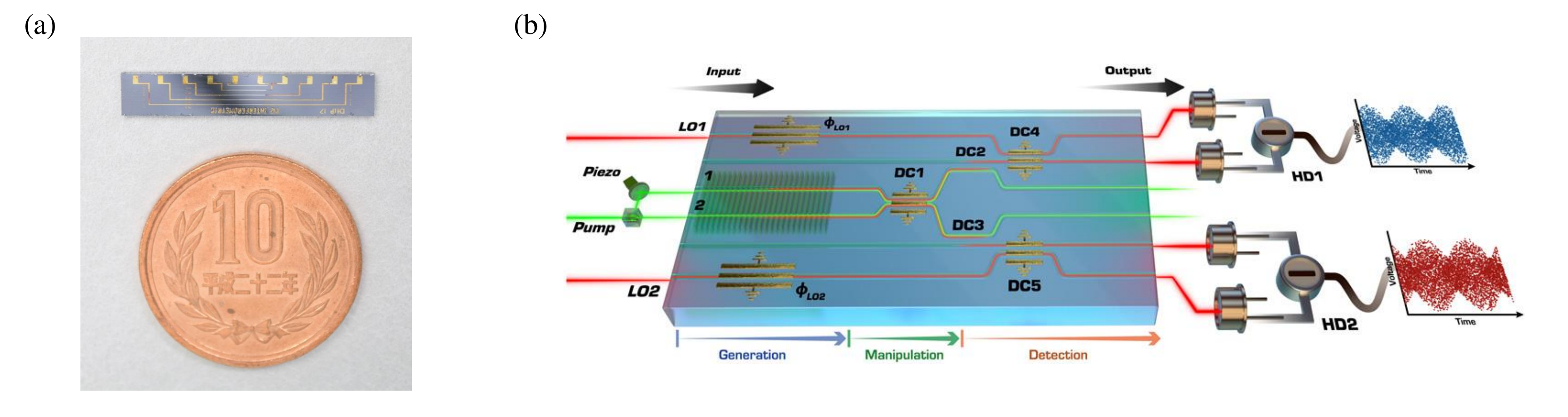}
\end{center}
\caption{Integrated photonic chips for CV quantum information processing.
(a) The silica-based photonic chip for generating CV entanglement in Ref.~\cite{15Masada}.
(b) The LN-based photonic chip for generating, mixing, and measuring squeezed beams. Reprinted from \cite{18Lenzini}. \copyright~The Authors, some rights reserved; exclusive licensee AAAS. Distributed under a CC BY-NC 4.0 license http://creativecommons.org/licenses/by-nc/4.0/.
}
\label{fig:IntegratedCircuits}
\end{figure*}

\subsection{Scaling up with integrated optics}\label{subsec:Integrated}

Another important research direction for scaling up optical QC
is to miniaturize optical circuits on small chips~\cite{20Wang2,20Elshaari}.
Traditional quantum optics experiments were based on bulky optical components separately sitting on a tabletop. In this case, the optical circuit is unavoidably large and susceptible to the instability of the surrounding environment, thus lacking scalability. In the same way as the evolution of on-chip integration of electronic components, there has been a tremendous effort in integrating and miniaturizing optical components on photonic chips. Photonics chips offer a promising path to packaging many quantum light sources, interferometers, and detectors compactly. There are many advantages for the on-chip integration, not to mention the miniaturization of large-scale optical circuits.
First, on-chip integration improves the stability of the optical circuits compared to the bulky setups that are susceptible to vibration and temperature change.
Second, it enhances the precision of quantum operations by avoiding spatial mode mismatching between propagating beams and reducing propagation losses with an appropriate choice of materials.
Third, optical circuits become reconfigurable and versatile because
on-chip beam splitter transmissivity and phase shifts can be externally controllable.
Finally, electronic circuits for measurement and feedforward operations can also be integrated into the same chip, enabling very fast and low latency operations.
There are several material options for photonic chips, each of which has advantages for realizing a practical integrated quantum photonics platform.
Quantum light sources for single-photon or squeezed states can be integrated on the chips if the material has sufficient second-order nonlinearity, converting one pump photon into two photons, or third-order nonlinearity, converting two pump photons into another two photons.

One of the most typical materials for the integration is fused silica (SiO$_2$)
due to its low propagation loss and good coupling efficiency to optical fibers.
Silica-based photonic chips integrating beam splitters and phase shifters
have been used for the early demonstrations in single-photon-based quantum information processing,
such as high-fidelity two-photon interference and two-qubit gates~\cite{08Politi},
Shor's factoring algorithm~\cite{09Politi}, multi-particle quantum walks~\cite{10Peruzzo},
and boson sampling~\cite{13Spring}.
The silica-based chips were later applied to CV quantum information processing,
such as CV entanglement generation~\cite{15Masada} (Fig.~\ref{fig:IntegratedCircuits}(a)) and measurement-induced squeezing operations~\cite{18Meinecke}.
However, the silica-based circuits are based on low-refractive-index-contrast  waveguides
whose low transverse confinement leads to relatively large chip size.
Even simple circuits require a chip size of $\sim10$ cm.
Therefore, the silica-based platform is not suitable for the fabrication of large complex photonic circuits.
In addition, silica has no second-order nonlinearity and also much lower third-order nonlinearity compared to silicon. Thus, silica is not suitable for integrating high-level squeezed light sources, although single-photon sources on silica chips have been reported~\cite{signorini2020chip}.

On the other hand, silicon-based platforms such as silicon (Si) and silicon nitride (Si$_3$N$_4$) offer a very high-refractive-index contrast as well as compatibility with existing foundary processes~\cite{18Wang,19Taballione}.
Its high-index contrast leads to not only high component density by reducing the circuit size but also high transverse confinement that is suitable for nonlinear processes.
These materials do not possess second-order nonlinearity but have sufficient third-order nonlinearity, enabling us to fabricate and integrate on-chip squeezed light sources.
There have been several CV experiments with silicon-based platforms. For example, Ref.~\cite{19Paesani} used a silicon chip to produce squeezed states by four-wave mixing and perform Gaussian boson sampling in a fixed configuration.
Silicon nitride ring resonators have also been demonstrated to
generate non-degenerate two-mode squeezed states~\cite{20Vaidya}
and degenerate single-mode squeezed states~\cite{20Cernansky,20Zhao,21Zhang} by four-wave mixing or self-phase modulation.
One of such sources were later integrated into a programmable photonic chip based on silicon nitride,
demonstrating Gaussian boson sampling and its related algorithms~\cite{21Arrazola}.

Another reasonable choice for the material is lithium niobate (LiNbO$_3$, LN)
since it has second-order nonlinearity for more efficient squeezed light generation~\cite{16Alibart,20Qi}.
In addition, its large electro-optic coefficient enables fast amplitude and phase modulation,
which is suitable for switching and feedforward operations in CVQC.
LN waveguides themselves have been investigated as single-pass squeezed light sources~\cite{20Takanashi, 20Kashiwazaki,07Yoshino,08Eto}.
Several experiments reported more advanced integrated photonic chips based on LN waveguide,
integrating periodically poled waveguide segments for squeezed light generation
and other segments for couplers and phase shifters~\cite{02Kanter,18Lenzini,19Mondain} (Fig.~\ref{fig:IntegratedCircuits}(b)).
These experiments used proton-exchanged LN waveguides having low index contrast.
In contrast, high index contrast LN waveguides can be fabricated
by a lithium-niobate-on-insulator (LNOI) platform, which offers much better scaling properties similar to silicon-based platforms.
Fabrication of LNOI waveguides for quantum state generation and manipulation has also been investigated~\cite{18Krasnokutska,20Qi,21Chen}.

Photonic QC on integrated chips has often been troubled with optical losses
caused by waveguide roughness, material absorption, and coupling efficiency at input and output ports.
These effects are expected to be reduced by suitable material choice and improved fabrication processes.
Ultimately, it is favorable to integrate all elements, including quantum light sources, interferometers, and detectors, on the same chip to avoid coupling losses.
In addition to integration of sources and interferometers described above, there has been much effort towards
integration of detectors, such as homodyne detectors~\cite{18Raffaelli,19Zhang,21Tasker},
superconducting nanowire single-photon detectors~\cite{11Sprengers,12Pernice},
and transition-edge sensors for photon number resolving detection~\cite{13Calkins,19Hopker}.
Integrated chips for time-multiplexed CVQC may use off-chip fiber delay lines,
so the development of structures for near-unity chip-fiber coupling efficiency is also desired.
On-chip integration of suitable CVQC architectures with broad operational bandwidth would be the ultimate form of photonic quantum computers, leading to scalable and universal QC at an ultra-high clock rate.

\section{Fault-tolerant quantum computation with CVs}
\subsection{Quantum error correction with CVs}
In the previous section, we have seen the progress to scale up photonic QC by several approaches, which enable us to increase the number of processable modes and gate steps.
Considering the errors during quantum gates, however, the errors are accumulated and propagated via quantum gate operations, which leads to the destruction of quantum information even if each of these errors is small. Thus, we must remove the errors to implement large-scale QC.
To solve the problem of errors during QC, the QEC for qubits~\cite{calderbank1996good} has been widely developed.
By using the QECC~such as Knill's ${\rm C}_4/{\rm C}_6$ code~\cite{knill2005quantum}, Steane's seven qubit code~\cite{steane1996error,Steane1997}, or surface codes~\cite{bravyi1998quantum,dennis2002topological} in a fault-tolerant manner, the errors can be substantially reduced if the error per quantum operation is below a constant threshold value~\cite{preskill1998reliable,knill1998resilient,aharonov2008fault}.
This is called a threshold theorem.

The QEC for CVs was introduced by Braunstein~\cite{braunstein1998error}, Lloyd and Slotine~\cite{lloyd1998analog}, where the standard qubit-based QECC such as Shor's nine qubit code~\cite{shor1995scheme} was applied to the quadrature eigenbasis $\{ \ket{s}_q\} _{s\in \mathbb{R}}$. The QECC by Braunstein has been demonstrated experimentally in Ref.~\cite{aoki2009quantum}, where Shor's nine qubit code was encoded and decoded by using squeezed light and linear optics.
Although it is worthwhile demonstrating the principle of the QECC for CVs, unfortunately, error models of these protocols do not correspond to the realistic errors, for example, photon loss during QC. 
Bosonic codes overcome this problem, as described in Sec.~\ref{subsec:bosonic}.

\begin{table*}[h]
\caption{A family of bosonic codes.} \label{family}
\renewcommand{\arraystretch}{1.0}
\begin{tabular}{|p{1.6cm}|p{6.51cm}|p{4.1cm}|p{3.3cm}|}
\hline
 &\hspace{5pt}Superposition of (squeezed) coherent states &\hspace{10pt}Fock state superposition&\hspace{12pt}Hybrid approach\\
\hline
\end{tabular}
\renewcommand{\arraystretch}{1.2}
\begin{tabular}{|p{1.6cm}|p{3.0cm}|p{3.1cm}|p{4.1cm}|p{3.3cm}|}
\hline
Example & \hspace{20pt}{Cat code}~\cite{ralph2003quantum}& \hspace{18pt}GKP qubit ~\cite{01Gottesman}& \hspace{18pt}Binomial code~\cite{michael2016new}& \hspace{2pt}Cat-photon qubit~\cite{lee2013near}\\ \hline
${\ket 0}_{\rm L}$ state 
&\centerline{$\ket{\alpha}$}   
&\centerline{$\sum\limits_{m=- \infty}^{\infty}\ket{2m\sqrt{\pi}}_{q}$ }
&\centerline{$\frac{1}{\sqrt{2^N}}\sum\limits_{p\in{\rm even}}^{[0,N+1]}
\sqrt{\binom{N+1}{p}}{\small \ket{p M}}$}
&\centerline{$\ket{+}\ket{\alpha}$}  \\ \hline
${\ket 1}_{\rm L}$ state
&\centerline{$\ket{-\alpha}$}
&\centerline{$\sum\limits_{m=- \infty}^{\infty}\ket{(2m+1)\sqrt{\pi}}_{q}$}
&\centerline{$\frac{1}{\sqrt{2^N}}\sum\limits_{p\in{\rm odd}}^{[0,N+1]}
\sqrt{\binom{N+1}{p}}{\small \ket{p M}}
 $}
& \centerline{$\ket{-}\ket{-\alpha}$}\\ \hline
Experiment (examples)
& Optical setup~\cite{etesse2015experimental}, Superconducting circuits~\cite{vlastakis2013deterministically}
& Ion trap system~\cite{fluhmann2019encoding}, Superconducting circuits~\cite{campagne2020quantum}.
&\leftline{Superconducting circuits}~\cite{hu2019quantum}
&Optical setup~\cite{jeong2014generation}\\ 
\hline
\end{tabular}

\begin{tabular}{|p{1.6cm}|p{6.51cm}|p{4.1cm}|p{3.3cm}|}
\hline
Other code examples
& Multicomponent cat code~\cite{bergmann2016quantum}, Rotation symmetric cat code~\cite{grimsmo2020quantum}, Pair cat code~\cite{albert2019pair}
& NOON code~\cite{bergmann2016quantum2, kok2002creation}, Loss tolerant code~\cite{chuang1997bosonic,Ralph2005,wasilewski2007protecting}
&\centerline{ N/A }\\
\hline
\end{tabular}
\renewcommand{\arraystretch}{1.0}
\end{table*}

\subsection{Bosonic codes}\label{subsec:bosonic}
\subsubsection{A family of bosonic codes}\label{subsubsec:family}
Bosonic codes encode discrete quantum information into bosonic modes. 
The bosonic code is a key ingredient to protect quantum information against errors from an environment such as photon loss. 
By concatenating bosonic codes with standard qubit-based QECCs such as Steane's seven qubit code~\cite{steane1996error,Steane1997}, one can implement large-scale QC thanks to the threshold theorem.
There are a variety of bosonic codes for encoding quantum information in a CV system~\cite{albert2018performance}, where they could be classified into three categories: a superposition of coherent states (or squeezed coherent states), a superposition of a finite number of Fock states, and a hybrid approach of them.
In this review, we classify the GKP qubit~\cite{01Gottesman} into the first category. We will explain more details of the GKP qubit in subsequent subsections.
Table~\ref{family} summarizes examples of the bosonic codes.

The first category includes the cat code, which is a superposition of a few coherent states~\cite{cochrane1999macroscopically,albert2019pair,niset2008experimentally}.
The logical 0 and 1 states for one of the simplest cat codes are defined as $\ket{0}_{\rm L}=\ket{\alpha}$ and $\ket{1}_{\rm L}=\ket{-\alpha}$, respectively, where $\ket{\alpha}$ is a coherent state with a real amplitude $\alpha$.
Since the overlap $\braket{-\alpha|\alpha}=e^{-2\alpha^2}$ is not zero, there is the probability of misidentifying the logical state, i.e., the bit-flip error. By using large $\alpha$, this probability becomes effectively zero~\cite{jeong2002efficient,lund2008fault,wickert2010entanglement}.
For a photon loss channel with the efficiency $\eta$, the probabilities of the bit- and phase-flip errors after photon loss are given by $e^{-2\eta\alpha^2}$ and $\frac{1}{2}(1-e^{-2(1-\eta)\alpha^2})$, respectively~\cite{wickert2010entanglement}.

The second category includes the binomial code composed of a superposition of Fock states with coefficients which obey a binomial distribution~\cite{michael2016new}.
The binomial code can correct amplitude damping, displacement, and dephasing errors by detecting the number parity of photons. The logical 0 and 1 states for the binomial code, which can correct $L$ photon losses, $G$ photon gain errors, and $D$ dephasing events, are given by
$\ket{0(1)}_{\rm L}=\frac{1}{\sqrt{2^N}}\sum_{p\in{\rm even(odd)}}^{[0,N+1]}\sqrt{\binom{N+1}{p}}{\small \ket{p M}}$, where $M=L+G+1$ and $N={\rm max} \{ L,G,2D\}$.

For the third category, there are several proposals for a hybrid approach of cat and photonic qubits~\cite{lee2013near,jeong2014generation,15Andersen,lee2015nearly,kapit2016hardware,lau2016universal,tiranov2016demonstration}.
The logical 0 and 1 states for a typical cat-photon qubit are given by $\ket{0}_{\rm L}=\ket{+}\ket{\alpha}$ and $\ket{1}_{\rm L}=\ket{-}\ket{-\alpha}$ with $\ket{\pm}=(\ket{H}\pm\ket{V})\sqrt{2}$, respectively, where $\ket{H}$ and $\ket{V}$ are horizontal and vertical polarization states of a photon, respectively.
The cat-photon qubit allows us to implement near-deterministic quantum teleportation~\cite{lee2013near}.
For large-scale QC with the hybrid approach, the architecture for fault-tolerant QC with the hybrid of photon and cat qubits has been proposed~\cite{omkar2020resource,omkar2021highly}.

\subsubsection{GKP qubit}\label{subsubsec:gkp}
Among bosonic codes, the GKP qubit proposed by Gottesman, Kitaev, and Preskill~\cite{01Gottesman} is the most promising alternative to photon-based qubits, which provides both universality and fault tolerance for CVQC.
A GKP qubit encodes a qubit into the oscillator's position and momentum quadratures.
Indeed, the GKP qubit can correct small random displacement errors for both $q$ and $p$ quadratures.
Furthermore, the GKP qubit has an excellent error tolerance against noises that frequently occur in an optical system, e.g., photon loss, compared with cat and binomial codes~\cite{albert2018performance}.  
GKP qubits inherit the advantage of squeezed vacuum states on optical implementation; they can be entangled by only beam-splitter coupling.
By virtue of this feature, the large-scale CV cluster state combined with the GKP qubits allows large-scale QC if these states have a sufficient squeezing level~\cite{menicucci2014fault}.
Experiments involving trapped ions~\cite{fluhmann2019encoding} and superconducting circuits~\cite{campagne2020quantum} have demonstrated the generation of the GKP qubit.
The squeezing level of the GKP qubit generated in Ref.~\cite{campagne2020quantum} is sufficient for fault tolerance~\cite{18Fukui,fukui2019high}.
In recent years, the GKP qubit has gathered a lot of interest for various applications.
In addition to the application for QC~\cite{fukui2018tracking,noh2020encoding, fukui2021efficient}, the GKP qubit performs well for quantum communication~\cite{Bennett1984,Briegel1998,Kimble2008} thanks to the robustness against photon loss~\cite{albert2018performance}. Indeed, recent results show that using GKP qubits may greatly enhance the distance of quantum communication~\cite{fukui2021all,rozpkedek2021quantum}.

\begin{figure}[b]
 \centering \includegraphics[angle=0, scale=1.0]{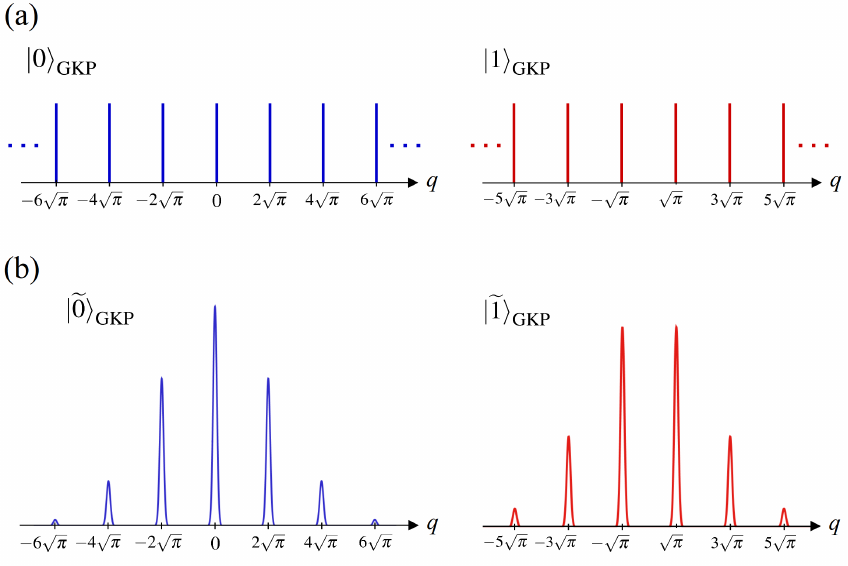} 
\caption{The codeword for the GKP qubit. (a) Ideal GKP qubit with infinite squeezing. (b) The GKP qubit with finite squeezing. }
\label{GKPqubits}
\end{figure}

The ideal code states of the GKP code are Dirac combs in $q$ and $p$ quadratures and are described as
\begin{eqnarray}
\ket {{0}}_{\rm GKP}= \sum_{m=- \infty}^{\infty}\ket{2m\sqrt{\pi}}_q,  \label{idealgkp0}\\
\ket {{1}}_{\rm GKP}= \sum_{m=- \infty}^{\infty}\ket{(2m+1)\sqrt{\pi}}_q,  \label{idealgkp1}
\end{eqnarray}
respectively, as shown in Fig.~\ref{GKPqubits}(a).
Since the ideal GKP code is not a normalizable state, physical states for the GKP code are finitely squeezed approximations to these and are often modeled as a comb of Gaussian peaks of variance $\delta^2$ with separation $2\sqrt{\pi}$ damped by a larger Gaussian envelope of variance$1/\kappa^2$. 
In the position basis, the logical 0 and 1 states of approximate code states, $\ket {\widetilde{0}}_{\rm GKP}$ and $\ket {\widetilde{1}}_{\rm GKP}$, are given by
\begin{eqnarray}
\ket {\widetilde{0}}_{\rm GKP} &\propto &  \sum_{m=- \infty}^{\infty} \int e^{-\frac{[(2m)\sqrt{\pi}]^2}{2(1/\kappa^2)}}e^{-\frac{(s-2m\sqrt{\pi})^2}{2\delta^2}}\ket{s}_q  ds,    \label{gkp0}\\ 
\ket {\widetilde{1}}_{\rm GKP} &\propto &  \sum_{m=- \infty}^{\infty} \int e^{-\frac{[(2m+1)\sqrt{\pi}]^2}{2(1/\kappa^2)}} 
e^{-\frac{[s-(2m+1)\sqrt{\pi}]^2}{2\delta^2}}\ket{s}_q  ds, \label{gkp1}     
\end{eqnarray}
respectively.
In the case of infinite squeezing ($\kappa \rightarrow 0$, $ \delta \rightarrow 0$) the states become ideal GKP qubits. 
In this review, we choose $\kappa^{2}$ and $\delta^{2}$ so that the variance of each peak in the position and momentum observables is equal to $\sigma^{2}_{\rm gkp}$, i.e., $\kappa^{2} = \delta^{2}=2\sigma^{2}_{\rm gkp}$.
These approximate states are not orthogonal, and there is a probability of misidentifying $\ket {\widetilde{0}}_{\rm GKP}$ as $\ket {\widetilde{1}}_{\rm GKP}$ (and vice versa) in the computational basis measurement.
The computational basis measurement is implemented by the measurement in the $q$ quadrature and binning to the integer multiple of $\sqrt{\pi}$. 
A qubit-level measurement error occurs when the measured outcome is more than $\sqrt{\pi}/2$ away from the correct outcome.
The probability of misidentifying the bit value of a non-ideal GKP qubit with the variance $\sigma_{\rm gkp}^2$,  $P_{\rm fail}(\sigma_{\rm gkp}^2)$, is approximately given by
\begin{equation}
P_{\rm fail}({\sigma_{\rm gkp}}^2) = 1-\int_{\frac{-\sqrt{\pi}}{2}}^{\frac{\sqrt{\pi}}{2}} dx \frac{1}{\sqrt{2\pi {\sigma_{\rm gkp}} ^2}} {\rm exp}\left(-\frac{x^2}{2{\sigma_{\rm gkp}} ^2}\right).
\label{gkperr}
\end{equation}

\subsubsection{Representation of the GKP qubit}
We review further studies for the mathematical formalism of the GKP qubit, which is an important research direction to understand CVQC with the GKP qubits. 
Here we will see three formalisms for the GKP qubits.

The first formalism is the normalization of the GKP qubit.
In addition to the state in Eqs.~(\ref{gkp0}) and (\ref{gkp1}), there are various ways to approximate the ideal states in Eqs.~(\ref{idealgkp0}) and (\ref{idealgkp1}) as physical states, i.e., normalizable states.
In Ref.~\cite{tzitrin2020progress}, Tzitrin {\it et al.}~summarized various ways to describe the GKP qubit as a physical state such as the following ways:
(1) The states in Eqs.~(\ref{gkp0}) and (\ref{gkp1}).
(2) The state suffered from a Fock damping operator $\hat{E}(\epsilon)=e^{-\epsilon\hat{n}}$ with $\epsilon>0$ and $\hat{n}=\hat{a}^{\dag}\hat{a}$, described by $\hat{E}(\epsilon) \ket {{0/1}}_{\rm GKP}$~\cite{menicucci2014fault}.
(3) The state suffered from an operator of coherent random shifts $\hat{G}=\sqrt{2/\pi \Delta^2}\int d^2\alpha e^{-|\alpha|^2/\Delta^2}\hat{D}(\alpha)$, defined by $\hat{G}\ket {{0/1}}_{\rm GKP}$.
The equivalence of the above ways (1)-(3) was analytically shown in Ref.~\cite{matsuura2020equivalence}.

The second formalism is the modular subsystem decomposition developed by Pantaleoni {\it et al.}~\cite{pantaleoni2020modular}, where arbitrary CV state can be decomposed into discrete variable and CV parts.
For a particular case of this formalism, the ideal GKP qubit can be described as $\ket{\phi}_{\rm GKP}=\ket{\phi}_{\mathcal{L}}\otimes \ket{+_{I}}_{\mathcal{G}}$, where $\ket{\phi}_{\mathcal{L}}=c_0\ket {{0}}_{{\rm GKP},{\mathcal{L}}}+c_1\ket {{1}}_{{\rm GKP},{\mathcal{L}}}$ $(|c_0|^2+|c_1|^2=1)$ represents the two-dimensional qubit subspace for the logical mode, and $\ket{+_{I}}_{\mathcal{G}}=\sum_{m=- \infty}^{\infty}\ket{m\sqrt{\pi}}_{q,{\mathcal{G}}}$ is represented as a state in the other infinite-dimensional space for the gauge mode~\cite{pantaleoni2020modular}.
Then, the normalizable GKP qubit can be described as $\hat{E}(\epsilon)\ket{\phi}_{\rm GKP}=\hat{E}(\epsilon)(\ket{\phi}_{\mathcal{L}}\otimes\ket{+_{I}}_{\mathcal{G}})$ by using the Fock damping operator $\hat{E}(\epsilon)$.
One of the applications of this technique is to define the Bloch sphere for the GKP qubit. 
Specifically, the density operator of the GKP qubit is obtained by tracing out the gauge mode as $\hat{\rho}(\epsilon)={\rm Tr}_{\mathcal{G}}[\hat{E}(\epsilon)\ket{\phi}_{\rm GKP}\bra{\phi}\hat{E}(\epsilon)]$. Then, $\hat{\rho}(\epsilon)$ provides information about the Bloch sphere for the qubit from $\hat{\rho}(\epsilon)=\frac{1}{2}\sum_{i=0}^{3}s_i\hat{\sigma}_i$, where $\hat{\sigma}_i$ and $s_i$ are the Pauli/identity matrix and the Stokes parameter~\cite{tzitrin2020progress}. 
Additionally, many studies have recently been conducted on the mathematical formalism using the modular subsystem decomposition~\cite{walshe2019robust,walshe2020continuous,mensen2020phase,tzitrin2020progress,
pantaleoni2021subsystem,pantaleoni2021hidden}.

The third formalism is the Wigner representation of the GKP qubit, which is also useful to characterize the GKP qubit.
As mentioned in the original GKP's paper~\cite{01Gottesman}, the Wigner function of the ideal GKP qubit, $\ket {{0/1}}_{\rm GKP}$, can be described by the sum of delta functions.
In Ref.~\cite{bourassa2021fast}, it has been shown that the Wigner function of the normalizable GKP qubit, $\hat{E}(\epsilon) \ket {{0/1}}_{\rm GKP}$, can be also described by the sum of Gaussian functions. This enables us to efficiently simulate the unitary time evolution of the GKP qubit.
Also, other non-Gaussian states such as the cat code and Fock states can be represented by this formalism.
Consequently, this formalism allows us to efficiently perform the numerical simulation of quantum information processing where several bosonic codes evolve simultaneously~\cite{killoran2019strawberry}.

\subsubsection{Universality for the GKP qubit}
To realize non-Gaussian gates and universality for fault-tolerant QC with GKP qubits, the cubic phase gate is expected to be a typical resource~\cite{01Gottesman}.
The cubic phase gate, however, may not be appropriate for the GKP qubit in terms of the gate fidelity~\cite{hastrup2020cubic}.
To achieve universality with the GKP qubit, the GKP magic state teleportation will be preferable compared with the cubic phase gate if we can prepare the GKP magic state with sufficient fidelity for the magic state distillation~\cite{bravyi2005universal,reichardt2005quantum,campbell2010bound}.
Ref.~\cite{konno2021non} has investigated the way to implement the GKP magic state teleportation in an optical setup.
For the preparation of the magic state of the GKP qubit, Baragiola {\it et al.} proposed an impressive scheme to prepare a noisy magic state from a logical GKP qubit and a vacuum state by only Gaussian operations and measurement.
Then, a high-fidelity magic state is prepared from many noisy magic states by using the magic state distillation~\cite{bravyi2005universal}. 
This implies that the preparation of the GKP logical basis and Gaussian operations allow us to implement universal and fault-tolerant QC.
In contrast, Ref.~\cite{yamasaki2020cost} proposed a more efficient scheme to achieve universal and fault-tolerant QC without the magic state distillation, where high-fidelity GKP magic states are directly prepared and combined with Gaussian operations.

\subsubsection{Error model for the GKP qubit}\label{subsubsec:error}
Here we will review an error model for the GKP qubit to lay the groundwork for the description of the QEC with the GKP qubit.
We describe the additive Gaussian noise (AGN)~\cite{holevo2001evaluating,01Gottesman,Harrington2001}, also known as a Gaussian random displacement channel, as a convenient error model.
The AGN is a kind of the Gaussian quantum channel (GQC) and is a common type of noise in bosonic systems.
The AGN randomly displaces the state in phase space according to a Gaussian distribution, which is described by the
superoperator $\mathcal G_\xi$ acting on density operator $\hat{\rho}$ as
\begin{eqnarray}
\label{eq:GQC}
\hat{\rho} \to \mathcal G_\xi (\hat{\rho}) &= \frac{1}{\pi{\xi }^2}\int d^2\alpha\, \mathrm{e}^{-| \alpha |^2/{{\xi}^2}}\hat{D}( \alpha ) \hat{\rho} \hat{D}^{\dagger }( \alpha ) ,
\end{eqnarray}
where $\hat{D}(\alpha) = e^{\alpha \hat{a}^\dag - \alpha^* \hat{a}}$ is the phase-space displacement operator. With $\alpha = (v + i u)/\sqrt{2}$, the position~$q$ and momentum~$p$ are displaced independently as $ \hat{q} \to \hat{q} + v$, $ \hat{p} \to \hat{p} + u $,
where $v$ and $u$ are real Gaussian random variables with mean zero and variance $\xi ^2$. Therefore, the AGN maintains the locations of the Gaussian peaks in the probability
for the measurement outcome, but it increases the variance of each peak by~$\xi ^2$ in both quadratures. 
Notice that the outputs of this channel are mixed states, even when $\hat{\rho}$ is pure.

The dominant noise channel in an optical setup is photon loss.
We will see that photon loss combined with the amplification techniques can be modeled by the increase of a variance, likewise the AGN.
The equivalence between the photon loss combined with amplifiers and the AGN can provide the way to apply the numerical results such as the threshold for fault-tolerant QC with the AGN to that with the photon-loss channel.

The photon-loss channel can be modeled as an unwanted beam-splitter coupling with an environmental vacuum state.
In the Heisenberg picture, the loss channel transforms the $\hat{q}$ and $\hat{p}$ quadratures as
\begin{equation}
\hat{q} \to   \sqrt{\eta}  \hat{q}+\sqrt{1-\eta}\hat{q}_{\rm vac}, \hspace{10pt} \hat{p} \to  \sqrt{\eta}  \hat{p}+\sqrt{1-\eta}\hat{p}_{\rm vac}, 
\end{equation}
where $\eta$ is the efficiency, and $\hat{q}_{\rm vac}(\hat{p}_{\rm vac})$ is the position (momentum) quadrature of the vacuum state.
After the loss channel, the variances of the GKP states in the $\hat{q}$ and $\hat{p}$ quadratures are transformed as
\begin{equation}
{\sigma^{2}} _{\rm gkp}   \to   {\eta}\sigma^{2}_{{\rm gkp}} +\frac{1-\eta}{2}.
\end{equation}

In Refs.~\cite{albert2018performance, noh2018quantum}, the equivalence between the AGN and the photon-loss channel combined with the amplification has been studied. 
At the cost of additional Gaussian noise, photon loss can be converted into a random displacement error channel, i.e., AGN by applying phase-insensitive amplifiers. 
The phase-insensitive amplifiers may be implemented either after or before the photon-loss channel, and these two methods are called post- and preamplification, respectively.
For the postamplification, the quadrature variances of the GKP qubits in both $\hat{q}$ and $\hat{p}$ are transformed as
$\sigma^2_{\rm gkp} \to   \sigma^2_{\rm gkp} + \frac{1-\eta}{\eta}$.
For the preamplification, the variances are transformed as $\sigma^2_{\rm gkp} \to \sigma^2_{\rm gkp} + 1-\eta.$
Note that the latter introduces less noise than the former. 
In addition to the phase-insensitive amplifiers, rescaling the outcome of the homodyne measurement is another method for the amplification.
Unlike the phase-insensitive amplifiers, the rescaling is implemented after the measurement of the target quantum state.
In the rescaling, the measurement outcome $m$ is rescaled on a classical computer as $m/\sqrt{\eta}$. The variances are transformed as $\sigma^2_{\rm gkp} \to  \sigma^2_{\rm gkp} + (1-\eta)/2\eta$. 
The performance of the above three techniques has been compared in the context of the quantum repeater protocol in Ref.~\cite{fukui2021all}.
Additionally, the rescaling technique has been applied to fault-tolerant QC with the GKP qubits~\cite{fukui2019high}.

\subsubsection{QEC with the GKP qubit} \label{subsubsec:sqec}
We review the way to correct small displacement errors in $q$ and $p$ quadratures~\cite{01Gottesman}, which is referred to as the single-qubit level QEC (SQEC) in the following. 
We consider errors derived from the variances of the input qubit and auxiliary states.

In the SQEC for the $q$ quadrature, a single ancilla qubit is entangled with the data qubit by the CX gate, where the ancilla qubit is the target qubit, as illustrated in Fig.~\ref{sqec}. The ancilla qubit is prepared in the state $\ket {\widetilde{+}}_{\rm GKP}$ to prevent us from identifying the bit value of the data qubit in the subsequent syndrome measurement. The CX gate, which corresponds to the operator $e^{-i\hat{q}_{\rm D}\hat{p}_{\rm A}}$ for CVs, transforms the quadratures as
\begin{eqnarray}
 \hat{q}_{\rm D} &\to &   \hat{q}_{\rm D}, \hspace{10pt}
 \hat{p}_{\rm D} \to  \hat{p}_{\rm D} - \hat{p}_{\rm A} ,  \\
 \hat{q}_{\rm A} &\to &  \hat{q}_{\rm A}+ \hat{q}_{\rm D}, \hspace{10pt}
 \hat{p}_{\rm A} \to   \hat{p}_{\rm A},
\end{eqnarray}
where $\hat{q}_{\rm D} (\hat{p}_{\rm D})$ and $\hat{q}_{\rm A} (\hat{p}_{\rm A})$ are the quadrature operators of the data and ancilla qubits in the $q$($p$) quadrature, respectively. 
Regarding the displacement errors in the $q$ and $p$ quadratures, the CX gate transforms the displacement errors as
\begin{eqnarray}
\overline{\Delta}_{\rm {\it q},{\rm D}} &\to & \overline{\Delta}_{\rm {\it q},{\rm D}} ,  \hspace{10pt}
 \overline{\Delta}_{\rm {\it p},{\rm D}} \to \overline{\Delta}_{\rm {\it p},{\rm D}}- \overline{\Delta}_{\rm {\it p},A}, \\
\overline{\Delta}_{\rm {\it q},A} &\to  &\overline{\Delta}_{\rm {\it q},A}+ \overline{\Delta}_{\rm {\it q},{\rm D}} , \hspace{10pt}
\overline{\Delta}_{\rm {\it p},A} \to \overline{\Delta}_{\rm {\it p},A},
\end{eqnarray}
where $\overline{\Delta}_{\rm {\it q},D} ( \overline{\Delta}_{\rm {\it p},D})$ and $\overline{\Delta}_{\rm {\it q},{\rm A}}  (\overline{\Delta}_{\rm {\it p},{\rm A}} )$ are the true values for the displacements of the data and ancilla qubits in $q$($p$), respectively.
We assume that the true values for the data (ancilla) qubit in the $q$ and $p$ quadratures obey the Gaussian distribution with the variance $\sigma^2_{{\rm D(A)},q}$ and $\sigma^2_{{\rm D(A)},p}$, respectively.
After the CX gate, we measure the ancilla qubit in the $q$ quadrature.
When $|\overline{\Delta}_{\rm {\it q},A}+ \overline{\Delta}_{\rm {\it q},D}|$ < $\sqrt{\pi}/2$, we obtain the measured displacement error of the ancilla qubit as ${\Delta}_{\rm m{\it q}, A}=\overline{\Delta}_{\rm {\it q},A}+ \overline{\Delta}_{\rm {\it q},D}$.
Then, we perform the displacement operation on the $q$ quadrature of the data qubit by $-{\Delta}_{\rm m{\it q}, A}$, and the true value of the displacement becomes $\overline{\Delta}_{\rm {\it q},D}-{\Delta}_{\rm m{\it q}, A}=-{\Delta}_{\rm {\it q}, A}$.
On other hand, when $|\overline{\Delta}_{\rm {\it q},A}+ \overline{\Delta}_{\rm {\it q},D}|$ > $\sqrt{\pi}/2$, we obtain the outcome of the displacement ${\Delta}_{\rm m{\it q}, A}=\overline{\Delta}_{\rm {\it q},A}+ \overline{\Delta}_{\rm {\it q},D}-\sqrt{\pi}$, assuming $\overline{\Delta}_{\rm {\it q},A}, \overline{\Delta}_{\rm {\it q},D}>0$.
\begin{figure}[t]
 \centering \includegraphics[angle=0, scale=0.9]{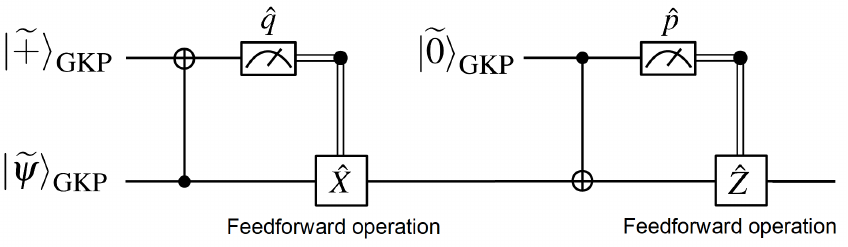} 
\caption{A quantum circuit for the single-qubit level QEC (SQEC) with the GKP qubit.}
\label{sqec}
\end{figure}
After the displacement operation by $-{\Delta}_{\rm m{\it q}, A}$, the true value for the data qubit in the $q$ quadrature becomes $\overline{\Delta}_{\rm {\it q},D}-{\Delta}_{\rm m{\it q}, A}=\overline{\Delta}_{\rm {\it q}, A}+\sqrt{\pi}$.
Since the displacement by $|\sqrt{\pi}|$ in the $q$ quadrature corresponds to the bit-flip error, the qubit-level error occurs on the data qubit in the $q$ quadrature if $|\overline{\Delta}_{\rm {\it q},A}+ \overline{\Delta}_{\rm {\it q},D}|$ > $\sqrt{\pi}/2$.
Assuming that the true value $\overline{\Delta}_{\rm {\it q},D(A)}$ obeys the Gaussian distribution with the variance $\sigma^2_{{\rm D(A)},q}$, the probability of the bit-flip error is given by $P_{\rm fail}(\sigma^2_{{\rm D},q}+{\sigma^2_{{\rm A},q}})$ using Eq.~(\ref{gkperr}).
Consequently, the SQEC in the $q$ quadrature can reduce the variance of the data qubit in the $q$ quadrature from $\sigma^2_{{\rm D},q}$ to ${\sigma^2_{{\rm A},q}}$ when $\sigma^2_{{\rm D},q}>{\sigma^2_{{\rm A},q}}$. 
For the variance of the data qubit in the $p$ quadrature, the initial variance $\sigma^2_{{\rm D},p}$ becomes $\sigma^2_{{\rm D},p}+{\sigma^2_{{\rm A},p}}$ after the SQEC in the $q$ quadrature. 

The SQEC in the $p$ quadrature can then be performed using the second ancilla qubit with the CX gate, where the ancilla is prepared in $\ket {\widetilde{0}}_{\rm GKP}$ and the data qubit is assumed to be the target qubit.
In a similar manner to the SQEC in $q$, but with the measurement of the ancilla in the $p$ quadrature, the SQEC for the data qubit in the $p$ quadrature can reduce the variance of the data qubit in the $p$ quadrature from $\sigma^2_{{\rm D},p}+{\sigma^2_{{\rm A},p}}$ to ${\sigma^2_{{\rm A2},p}}$, where ${\sigma^2_{{\rm A2},p(q)}}$ is the variance in the $p(q)$ quadrature for the second ancilla qubit. 
In the SQEC in $p$, there is the phase-flip error whose probability is given by $P_{\rm fail}(\sigma^2_{{\rm D},p}+{\sigma^2_{{\rm A},p}}+{\sigma^2_{{\rm A2},p}})$ using Eq.~(\ref{gkperr}). 

As a result of the sequential SQECs, the variances of the data qubit are replaced by those of the ancilla qubits, as $(\sigma^2_{{\rm D},q}, \sigma^2_{{\rm D},p}) \to ({\sigma^2_{{\rm A},q}}+{\sigma^2_{{\rm A2},q}},{\sigma^2_{{\rm A2},p}})$.
This kind of variance reduction with the SQEC is helpful for the QEC against the AGN which increases the variance, if we can prepare ancilla qubits whose variance is sufficiently small.
We note that the SQEC can be applied to a photon-loss channel by transforming the channel into an AGN with the amplification technique as described in Sec.~\ref{subsubsec:error}.
While we have used the CX gate for the SQEC, the SQEC can be also performed by using a beam-splitter coupling and a squeezing operation instead of the CX gate. The error analysis using such a setup was studied in Refs.~\cite{Glancy2006,wan2020memory}.

\subsection{Fault-tolerant QC with the GKP qubits}
\subsubsection{Higher-level encoding for GKP qubits}
Although the SQEC works well for the small displacement error, we need to use the logical-qubit level QEC to correct the displacement error greater than $\sqrt{\pi}/2$. 
In general, the error probability for each logical-qubit level operation should be below $10^{-12}-10^{-15}$~\cite{jones2012layered} to employ useful quantum algorithms.
Therefore, when we use GKP qubits for such algorithms, we must sufficiently reduce the logical qubit level errors
by concatenating the GKP qubits with the higher-level encoding, i.e., the qubit-based standard QECCs such as Steane's seven qubit code~\cite{steane1996error,Steane1997}, Knill's $C_{4}/C_{6}$ code~\cite{knill2005quantum}, the surface code~\cite{bravyi1998quantum,dennis2002topological}, and so on.
Then, the logical-qubit level error probability of the QECC can be suppressed to an arbitrary value if the physical-qubit level error is smaller than the threshold value.
In the case of the GKP qubits, the threshold value corresponds to the required initial squeezing level of the resource state, which determines the physical-qubit level error probability of the GKP qubits.

\begin{figure}[b]
 \centering\includegraphics[angle=0, width=1.0\columnwidth]{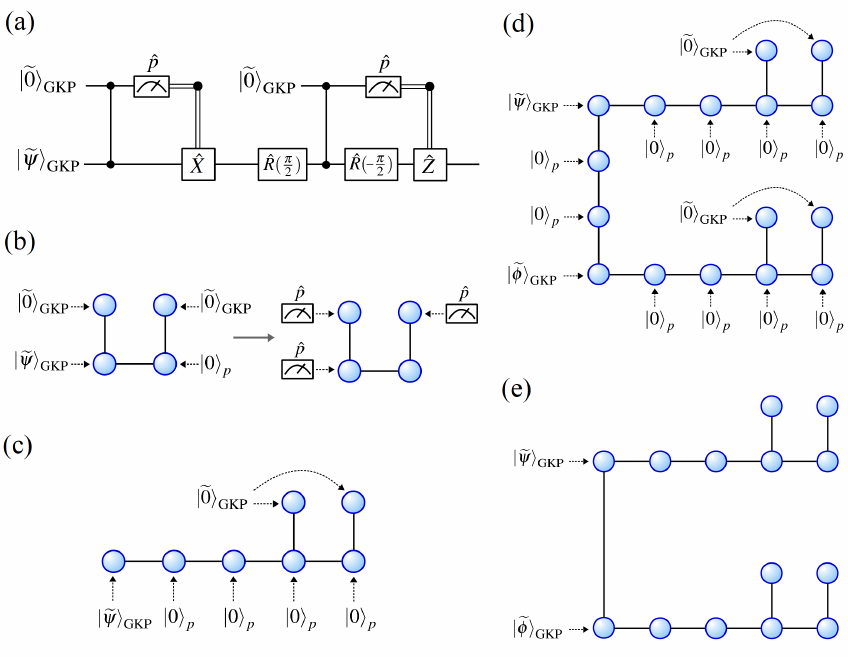}
 \caption{One-way QC based on the CV cluster state with the GKP qubit.
(a) A quantum circuit for the SEQC in the $q$ and $p$ quadratures, which corresponds to Fig.~\ref{sqec} (up to Fourier gates). 
(b) Cluster state representation corresponding to (a).
(c) A quantum circuit for one-way QC to implement an arbitrary single-mode quantum gate with the SQECs in the $q$ and $p$ quadratures. 
(d) Cluster state representation of the cluster state to implement the CZ gate with the SQECs in the $q$ and $p$ quadratures.
(e) Cluster state representation after the CZ gate between the data qubits, $\ket{\widetilde{\psi }}_{\rm GKP}$ and $\ket {\widetilde{\phi }}_{\rm GKP}$, via the measurements of the two ancilla qubits in the $p$ quadrature~\cite{menicucci2014fault}.
}
 \label{gkpcz}
\end{figure}

\subsubsection{Threshold for fault-tolerant QC}
A squeezing level is commonly used to represent the threshold for fault-tolerant CVQC, where a squeezing level $s$ is given by $s=-10{\rm log}_{10}2\sigma^2_{\rm gkp}$. 
Now, we move on to the threshold value for fault-tolerant CVQC with the GKP qubits. 
The first work for the threshold of the squeezing level was provided by Menicucci in 2014~\cite{menicucci2014fault}.
Also, Ref.~\cite{menicucci2014fault} is the first work that showed fault-tolerant QC is possible within the framework of one-way QC using large-scale CV cluster states. 
In the case of one-way CVQC, besides errors from an environment such as photon loss, errors derived from the CV cluster state itself accumulate during QC~\cite{menicucci2014fault}.
The GKP qubit overcomes this problem by performing the SQEC with CV cluster states, and qubit-level Clifford gates on the GKP qubits are implemented simultaneously by homodyne measurements on CV cluster states.
In the following, we see the implementation of the SQEC on the CV cluster state and the procedure to obtain the threshold of the squeezing level.

First, we see the way to perform the SQEC during one-way CVQC.
Figures~\ref{gkpcz}(a) and (b) show a quantum circuit and the cluster state representation for the SQECs, respectively, where Fig.~\ref{gkpcz}(a) corresponds to the quantum circuit in Fig.~\ref{sqec} up to Fourier gates.
Figure~\ref{gkpcz}(c) shows the cluster state to perform a single-mode quantum gate and the SQECs in the $q$ and $p$ quadratures. 
For the implementation of a single-mode quantum gate, the first four modes in the transverse direction are used (we recall that an arbitrary single-mode quantum gate can be implemented by the four-mode linear cluster state).
The SQECs in $q$ and $p$ quadratures are implemented by using ancillary GKP qubits entangled with the last two modes in the transverse direction, as can be understood from the correspondence between Figs.~\ref{gkpcz}(b) and (c). Therefore, the cluster state in Fig.~\ref{gkpcz}(c) enables us to perform the SQECs in $q$ and $p$ quadratures during single-mode one-way QC.
We note that there is the failure probability of SQECs, and the error probability for the single mode gate is given by the probability that at least one of the two SQECs fails.

In general, the CZ or CX gate sets the threshold for qubit-based fault-tolerant QC since the two-qubit gate is the noisiest gate among the universal gate set for qubits. 
Figures~\ref{gkpcz}(d) and (e) show the CZ gate between the data GKP qubits, $\ket{\widetilde{\psi }}_{\rm GKP}$ and $\ket {\widetilde{\phi }}_{\rm GKP}$, by the measurement of the squeezed vacuum states between the data qubits in the $p$ quadrature.
In this model~\cite{menicucci2014fault}, the CZ gate is followed by four SQECs, and thus the error probability for the CZ gate is given by the probability that at least one of the four SQECs fails.
We assume that the initial variance $\sigma^2_{\rm gkp}$ for the GKP qubit is the same as that of the initial momentum-squeezed vacuum states used to make the CV cluster state with a finite squeezing parameter $r$, $\sigma^2_{\rm in}=\frac{1}{2}e^{-2r}$.
Then, the Gaussian-distributed shift errors in two of the four SQECs (including noise of ancillary squeezed vacuum states after CVQC) have variance $7\sigma^2_{\rm in}$, and two others have variance $5\sigma^2_{\rm in}$~\cite{menicucci2014fault}.
Therefore, the probability that at least one of those corrections fails is $p_{\rm err}(\sigma^2_{\rm in})= 1-[1-P_{\rm fail}(7\sigma^2_{\rm in})]^2[1-P_{\rm fail}(5\sigma^2_{\rm in})]^2$.
Then, it can be expected that fault-tolerant QC is possible if $p_{\rm err}$ is smaller than the threshold value $p_{\rm FT}$ of the qubit-based QECCs~\cite{calderbank1996good,steane1996error,Steane1997,knill2005quantum}.
The threshold value $p_{\rm FT}$ depends on QECCs for qubits, e.g. $10^{-6}$--$10^{-2}$ in Refs.~\cite{preskill1998reliable,knill1998resilient,bravyi1998quantum,
dennis2002topological,knill2005quantum,fujii2010cluster}.
The threshold of the squeezing level can be obtained from the condition $p_{\rm FT}=p_{\rm err}(\sigma^2_{\rm in})$.
For example, when we use $p_{\rm FT}=10^{-6}$ for Steane's seven qubit code~\cite{Steane1997}, fault-tolerant QC can be realized if $\sigma^2_{\rm in}$ < 4.44 $\times$ $10^{-3}$. 
In this case, we obtain the threshold of the squeezing level as $-10{\rm log}_{10}2\sigma^2_{\rm in}\sim20.5$ dB.
The GKP qubit enables us to implement large-scale QC, but it is still challenging to experimentally generate the GKP qubit with the squeezing level 15.6--20.5 dB for qubit-based QECCs~\cite{menicucci2014fault}. 
In Secs.~\ref{subsubsec:analog},~\ref{subsubsec:hrm}, and~\ref{subsubsec:architecture}, we will see several efforts to reduce the threshold of the squeezing level.

\begin{figure}[t]
 \centering \includegraphics[angle=0, scale=0.95]{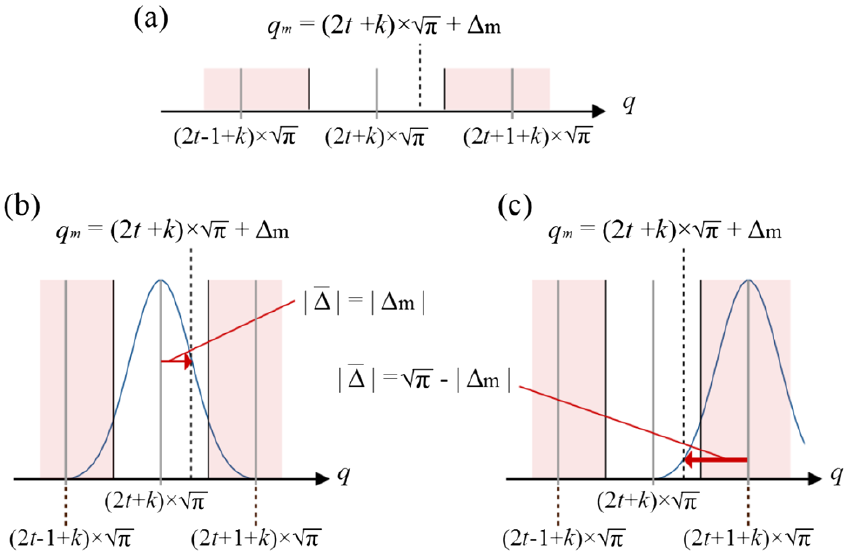} 
\caption{Analog quantum error correction. Reproduced from~\cite{fukui2017analog}.}
\label{aqec}
\end{figure}

\subsubsection{Analog QEC with GKP qubits} \label{subsubsec:analog}
The effect of noise on CV states is observed as the displacement error in an analog measurement outcome, which includes beneficial information for the QEC.
Thus, the analog measurement outcome is expected to improve the QEC performance compared to the QEC based on only binary information.
Analog QEC~\cite{fukui2017analog} provides such an improvement of the QEC performance by using the real-valued syndrome of a GKP qubit.
In the measurement of the GKP qubit, we obtain the measurement outcome $(2t+k)\sqrt{\pi}+\Delta_{\rm m}$, as shown in Fig.~\ref{aqec}(a), where $t$, $k$, and $\Delta_{\rm m}$ are integer, the bit value, and the measured displacement error.
As described in Sec.~\ref{subsubsec:sqec}, there are two possible events: one is no error, i.e., the true value for the displacement error $\overline{\Delta}$ is equal to $\Delta_{\rm m}$, as shown in Fig.~\ref{aqec}(b).
The other is the bit(phase)-flip error in the $q(p)$ quadrature, i.e., $|\overline{\Delta}|$ is equal to $\sqrt{\pi}-|\Delta_{\rm m}|$, as shown in Fig.~\ref{aqec}(c).
Then, we obtain two likelihoods for the two events as $f(\overline{\Delta})=1/ \sqrt{2\pi \sigma^2} e^{-\overline{\Delta}^2/(2\sigma^2)}$ since $\overline{\Delta}$ obeys the Gaussian distribution $f(\overline{\Delta})$.
The analog QEC applies these likelihoods to higher-level encoding, namely the qubit-based standard QECC.
Then, the most likely error pattern is selected to improve the performance of the decoding in the QEC.
For example, the likelihood is used for procedures of the decoding, such as a belief propagation for the concatenated code~\cite{poulin2006optimal,goto2013fault} and a minimum-weight perfect-matching algorithm for the surface code ~\cite{bravyi1998quantum,dennis2002topological,fowler2012surface}.

One may wonder how the analog QEC improves the QEC performance.
We focus on the attainable rate against the AGN with the code capacity model. 
We note that the attainable rate does not correspond to the threshold for fault-tolerant QC, while we see it as a simple example of the analog QEC. 
In the code capacity model, the noise is parametrized by a single variance $\sigma^2$ of the AGN, where the GKP qubits with zero variance (i.e., infinite squeezing) become those with the variance $\sigma^2$ after the AGN. 
The attainable rate for the GQC without the analog QEC can be obtained by the quantum capacities of the binary channel~\cite{calderbank1996good,lloyd1997capacity}, assuming that bit- and phase-flip errors occur with the probability $p_{\rm e}$ independently.
Using the binary Calderbank-Shor-Steane code~\cite{calderbank1996good,steane1996error}, the attainable rate $R$ is defined by $R > 1-2H_{2}(p_{\rm e})=1+2p_{\rm e}{\rm log}_{2}p_{\rm e}+2(1-p_{\rm e}){\rm log}_{2}(1-p_{\rm e})$, i.e., the rate is nonzero when $p_{\rm e}<0.11$, where $H_{2}(p_{\rm e})$ is the binary entropy function.
For the case of the GKP qubit, the error probability $p_{\rm e}$ is equal to $P_{\rm fail}(\sigma^2)$ described in Eq.~(\ref{gkperr}).
Since $P_{\rm fail}(\sigma^2)$ with $\sigma~\sim0.555$ corresponds to $p_{\rm e}\sim 0.11$, 
the nonvanishing quantum capacity for the AGN is defined when $\sigma<0.555$.
In Refs.~\cite{fukui2017analog,18Fukui}, it has been shown the analog QEC with Knill's ${\rm C}_4/{\rm C}_6$ code~\cite{knill2005quantum} and the surface code~\cite{bravyi1998quantum,dennis2002topological} can achieve $\sigma~\sim0.607$, which is larger than~$\sim0.555$.
This value reaches the hashing bound of the GQC, also known as a lower bound on the quantum capacity of the GQC~\cite{01Gottesman,Harrington2001}.
Thus, the analog QEC considerably improves the error tolerance, indicating the potential to improve the threshold for fault-tolerant QC.

\subsubsection{The highly reliable measurement}\label{subsubsec:hrm}
Another technique to improve the threshold was proposed in Ref.~\cite{18Fukui}, where the postselected measurement, referred to as the highly reliable measurement (HRM), is used to prevent the squeezing level from decreasing during the cluster state preparation.
Figures~\ref{hrm}(a) and (b) show the schematic of the HRM.
Here we see one of the peaks for the logical 0 state, which lies in position $2k\sqrt{\pi}$ with integer $k$ in the $q$ quadrature.
In Fig.~\ref{hrm}(a), the plain blue region and the red region with vertical lines represent the different code words of the logical 0 and 1 states, respectively.
In the conventional measurement, the red regions marked with vertical lines correspond to the probability of incorrect decision of the bit value. 
In the HRM, we set the dotted lines which represent an upper limit $\zeta$, as shown in Fig.~\ref{hrm}(b).
The yellow areas with horizontal lines show the probability that the results of the measurement are discarded by introducing $\zeta$. The vertical line areas show the probability that our method fails. 
By using the HRM, the probability of misidentifying the bit value, namely bit- and phase-flip errors, decreases with increasing $\zeta$ at the cost of reducing the success probability of the measurement.

\begin{figure}[t]
\centering \includegraphics[angle=0, scale=0.9]{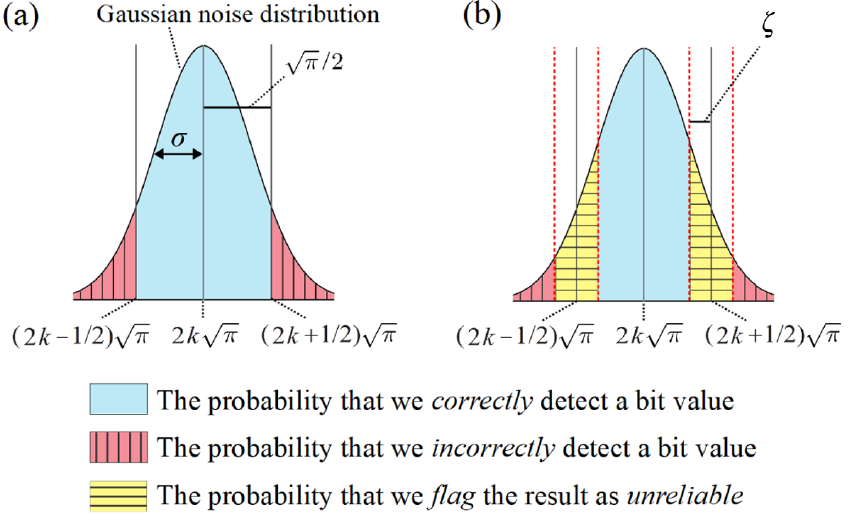} 
     \setlength\abovecaptionskip{10pt}
\caption{(a) The conventional measurement of the GKP qubit, where the Gaussian probability distribution following the displacement error of the GKP qubit has variance $\sigma^{2}$. (b) The highly-reliable measurement (HRM). Reproduced from~\cite{18Fukui}.}
\label{hrm}
\end{figure}

\begin{figure}[t]
\centering \includegraphics[angle=0, scale=0.95]{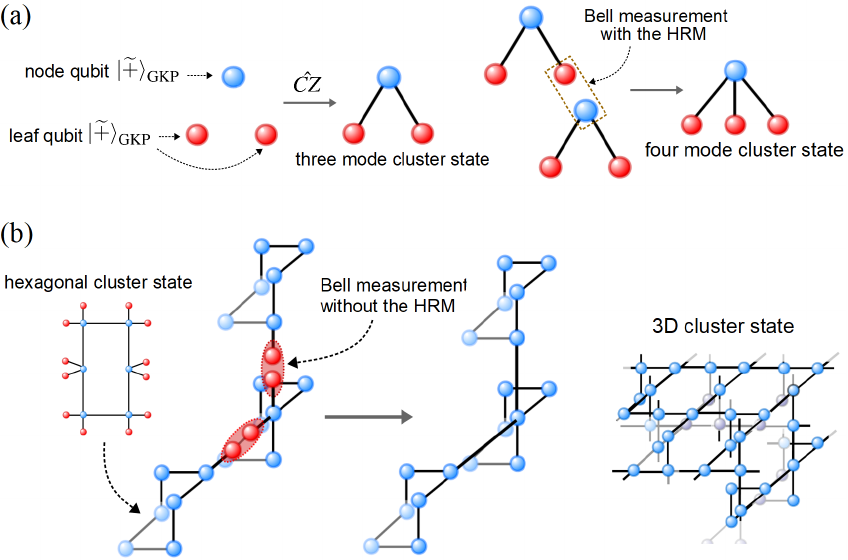} 
     \setlength\abovecaptionskip{10pt}
\caption{(a) The small-scale cluster state construction by using the Bell measurement with the HRM. 
(b) The three-dimensional (3D) cluster construction from the small-scale cluster state, which is referred to as the hexagonal cluster state, by using the Bell measurement without the HRM~\cite{18Fukui}.}
\label{3dconst}
\end{figure}

One of the applications of the HRM is the reliable construction of large-scale cluster states. 
Figures~\ref{3dconst}(a) and (b) show the schematic of the three-dimensional (3D) cluster state construction with the HRM, where the large-scale 3D cluster state can be used for topologically protected one-way QC with the surface code~\cite{raussendorf2007fault,raussendorf2007topological}. 
In Ref.~\cite{18Fukui}, the three mode cluster state, composed of a node qubit and two leaf qubits, is firstly prepared by the CZ gate, as shown in Fig.~\ref{3dconst}(a).
Then, the elementary cluster state, referred to as the hexagonal cluster state, is prepared from the three mode cluster states by using the Bell measurement with the HRM, where the HRM reduces the probability of misidentifying the bit values in the Bell measurement.
The reason for using the Bell measurement is that the entanglement generation using the Bell measurement does not increase the variance of the node qubit, while that using the CZ gate increases the variances due to the propagation of the displacement errors between qubits. 
Once the hexagonal cluster states are prepared probabilistically, the large-scale 3D cluster state is generated deterministically by using the Bell measurement without the HRM, as shown in Fig.~\ref{3dconst}(b).
In Ref.~\cite{18Fukui}, the fault-tolerant threshold of the squeezing level can be relaxed to less than 10 dB by combining the reliable 3D cluster state construction and the analog QEC. 
Fault-tolerant QC employing the HRM has been studied in several works~\cite{fukui2019high,seshadreesan2021coherent}.
Furthermore, the Gauss-Markov theorem in statistics is useful to improve the threshold introduced in Refs.~\cite{fukui2019high,yamasaki2020polylog,seshadreesan2021coherent}.
In addition, the HRM has been applied to loss-tolerant QECCs with the GKP qubits, e.g., Varnava's code~\cite{Varnava2006} to achieve long-distance quantum communication~\cite{fukui2021all} and the quantum parity code~\cite{Ralph2005} to realize the efficient decoding of the QEC~\cite{fukui2021efficient}.

\subsubsection{Architecture for fault-tolerant QC with the GKP qubits}\label{subsubsec:architecture} 
Promising architectures for fault-tolerant QC with the GKP qubit have been proposed recently in a superconducting circuit~\cite{20Terhal,noh2020fault,grimsmo2021quantum} and an optical setup~\cite{21Bourassa,tzitrin2021fault,21Larsen}.
In an optical setup, there are roughly two types of architectures for implementing large-scale QC.
The first one is the time-multiplexing approach (see for example Refs.~\cite{larsen2020architecture,21Larsen}), which is described in Sec.~\ref{subsubsec:time-one-way}.
The second one is the integrated approach (see Refs.~\cite{21Bourassa,tzitrin2021fault} as an example).
In the following, we see these two approaches. 

In the first time-multiplexing approach, a time-multiplexed cluster state for the resource of one-way QC is prepared by using a compact experimental setup. 
In Ref.~\cite{larsen2020architecture}, the noise analysis of the quantum gate and the threshold for fault-tolerant QC have been studied under an experimentally feasible noise model, employing an efficient way to perform one-way QC, called the macro node protocol~\cite{alexander2016flexible, 18Alexander}.
In Ref.~\cite{21Larsen}, the generation scheme of the 3D cluster state has been proposed by using time-multiplexing, where topologically protected QEC using the surface code on the 3D cluster state was considered. The threshold in this architecture was found to be 13.2 dB.
Experimental realization of one-way QC based on the time-multiplexing approach is described in Sec.~\ref{subsubsec:time-one-way}.

In the second integrated approach in Refs.~\cite{21Bourassa,tzitrin2021fault}, the 3D cluster state is generated from integrated photonic devices.
This architecture circumvents one disadvantage of the first approach, namely lossy long optical delay lines required for increasing the number of qubits, while the size of photonic devices becomes larger as the number of the logical qubits and the code distance of a surface code increase.
Ref.~\cite{21Bourassa} is the first work considering the probabilistic nature of the GKP qubit preparation schemes in one-way QC with GKP qubits. This probabilistic nature has been a crucial problem in large-scale QC.
In Ref.~\cite{21Bourassa}, the node for the event of the failure of the GKP qubit preparation is replaced by the node of the squeezed vacuum state. Then, one-way QC is performed on the 3D cluster state whose nodes are partially replaced by the squeezed vacuum states. 
This scheme allows us to implement large-scale QC even with non-deterministic GKP qubit preparation. On the other hand, it requires an optical switches for reconfigurability and inline squeezing for implementing the CZ gate.

These experimental requirements are removed by the scheme in Ref.~\cite{tzitrin2021fault}. Additionally, Ref.~\cite{tzitrin2021fault} proposed to construct the 3D cluster states from so-called qunaught states, which are also useful for quantum sensing applications~\cite{duivenvoorden2017single}.
The logical 0 and 1 states for the ideal qunaught state are given by $\ket {{0}}_{\rm naught}= \sum_{m=- \infty}^{\infty}\ket{2m\sqrt{2\pi}}_q$ and $\ket {{1}}_{\rm naught}= \sum_{m=- \infty}^{\infty}\ket{(2m+1)\sqrt{2\pi}}_q$, respectively.
For the 3D cluster state construction, an entangled pair of the GKP qubits is prepared from two approximate qunaught states with $\sigma_{\rm gkp}^2$ by using a beam-splitter coupling, where the variance of the GKP qubit composing the entangled pair is ($\sigma_{\rm gkp}^2,\sigma_{\rm gkp}^2$) in $q$ and $p$ quadratures. 
In the case of the entangled pair preparation by two GKP qubits with $\sigma_{\rm gkp}^2$ and the CZ gate, the variance of the GKP qubit is ($\sigma_{\rm gkp}^2,2\sigma_{\rm gkp}^2$). Thus, the qunaught state reduces the noise in the entangled pair, which is expected to improve the threshold of the squeezing level.

\subsubsection{Preparation of the GKP qubit in an optical setup}
In superconducting circuit and ion trap systems, the GKP qubit has been realized recently~\cite{fluhmann2019encoding,campagne2020quantum}, and the efficient preparation of the GKP qubit has been extensively studied~\cite{terhal2016encoding,weigand2018breeding,arrazola2019machine,eaton2019non,sabapathy2019production,
tzitrin2020progress,fukui2021efficient2}.
In an optical setup, to the best of our knowledge, the GKP qubit has not been realized, although there are several approaches such as the breeding protocol~\cite{vasconcelos2010all,weigand2018breeding}, the use of photon-number resolving detectors~\cite{su2019conversion,fukui2021efficient2}, the use of the interaction between light and matter qubits~\cite{pirandola2006continuous,motes2017encoding, hastrup2021generation}, the use of the cross-Kerr interaction~\cite{pirandola2004constructing,fukui2021generating}, and so on. 
Here we see the first two approaches, which can prepare the GKP qubit in an all-optical setup.

The breeding protocol proposed by Vasconcelos {\it et al.}~\cite{vasconcelos2010all} is the method to generate the GKP qubit by linear optics from squeezed cat states, where a cat state is defined as a superposition of two coherent states with opposite phases.
In the original protocol~\cite{vasconcelos2010all}, two squeezed cat states are interfered by a beam splitter and the homodyne measurement is performed on one of the modes in the $p$ quadrature, as shown in Fig.~\ref{prepgkp}(a).
When the measurement outcome is close to zero, the first step of the protocol succeeds.
Assuming that the wave function of the cat state has two Gaussian peaks illustrated in Fig.~\ref{prepgkp}(a), the output state has three Gaussian peaks.
Then, by repeating the breeding, the number of the Gaussian peaks grows, and the output states after each of the steps become similar to the GKP qubit.
The total success probability of the GKP qubit generation for the original protocol is too low due to the postselection in the homodyne measurement.
Ref.~\cite{terhal2016encoding} overcomes this problem by applying a phase estimation, where the postselection in the measurement is not required, but only the specific feedforward operation for phases depending on measurement outcomes is required. 
The scheme in Ref.~\cite{terhal2016encoding} is intended for the GKP qubit generation in microwave modes in superconducting circuits.
In Ref.~\cite{weigand2018breeding}, the breeding protocol with a phase estimation was developed by using linear optics. 
The breeding protocol has been experimentally demonstrated in Ref.~\cite{sychev2017enlargement}, while the scheme in Ref.~\cite{sychev2017enlargement} was not aimed to increase the number of Gaussian peaks, but to enlarge the amplitude of the cat state.

\begin{figure}[t]
 \centering \includegraphics[angle=0, scale=0.9]{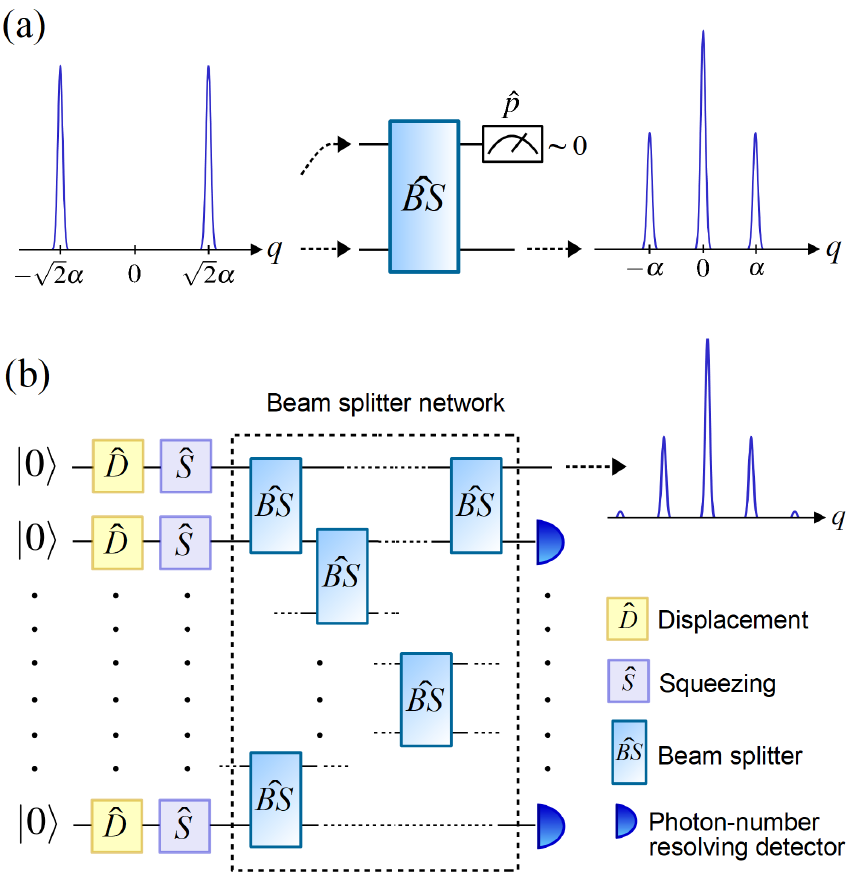} 
\caption{Preparation of the GKP qubit in an optical setup. (a) Breeding protocol~\cite{vasconcelos2010all}. (b) Optical quantum state synthesizer referred to as the Gaussian boson sampling (GBS)-based device~\cite{su2019conversion,tzitrin2020progress}.}
\label{prepgkp}
\end{figure}

Another method is to use the optical quantum state synthesizer consisting of linear optics and photon-number resolving detectors, which is referred to as the Gaussian boson sampling (GBS)-based device~\cite{su2019conversion,tzitrin2020progress}.
Figure \ref{prepgkp}(b) shows the schematic diagram for the optical quantum state synthesizer, where $l$ input vacuum states are initially displaced and squeezed, then interacted by a beam-splitter network, and finally those except for the output state are measured by photon-number resolving detectors. 
Depending on the pattern of the detected photon numbers and circuit parameters for linear optics, the output state is prepared as
$
\ket{\psi}_{\rm out}\approx \hat{U}\sum_{i=0}^{n_{\rm max}}\frac{c_{i}}{N}\ket{i}, \label{eqfock}
$
where $n_{\rm max}$, $c_{i}$, and $N$, are the truncated photon number in the Fock basis, the coefficient of the Fock state $\ket{i}$, and a normalization factor, respectively, and $\hat{U}$ is composed of single-mode Gaussian operations (e.g., squeezing, displacement, and rotation)~\cite{su2019conversion}.
The truncated photon number is given by $n_{\rm max}=\sum_{i=2}^{l}{m_{i}}$, where ${m_{i}}$ is the detected photon number in the $i$-th detector.
The coefficients $c_{i}$ depend on $m_{i}$ and the circuit parameters such as transmissivities of the beam splitters and the amount of the squeezing and displacement operations~\cite{su2019conversion}.
Those parameters are optimized by an optimization algorithm running on a classical computer so that the output state $\ket{\psi}_{\rm out}$ becomes close to the target non-Gaussian state. 
To prepare an arbitrary non-Gaussian state with $n_{\rm max}$, we need to optimize coefficients of the generated state up to $n_{\rm max}$ independently. 
The number of independent parameters, which we can access to optimize them, has been conjectured as ${(l+2)(l-1)}/{2}$~\cite{su2019conversion} and scales polynomially with $l$.
Thus, the GBS-based device with $l$ input states is expected to prepare an arbitrary state up to $\ket{n_{\rm max}}$ with $n_{\rm max}=(l+2)(l-1)/2-1$.

For fault-tolerant QC, the GBS-based device needs to generate well-approximated GKP qubits, which require sufficiently large $n_{\rm max}$.
However, the complexity for the classical numerical simulation of the GBS-based device scales exponentially with $l$.
This limits the number of coefficients we can optimize, considering $n_{\rm max}$ scales polynomially with $l$.
The complexity mainly comes from the calculation of a $loop$ $hafnian$ which is contained in the class of \#P-complete problems~\cite{valiant1979complexity,quesada2019franck}.
In fact, the complexity of the calculation of a hafnian has been used for achieving quantum supremacy over a classical computer by a protocol called GBS~\cite{20Zhong,aaronson2011computational}.
Recently, the efficient backcasting search has been developed to solve the problem of time complexity for arbitrary non-Gaussian state generation~\cite{fukui2021efficient2}. 
Also, Ref.~\cite{fukui2021efficient2} has shown a specific configuration of an optical circuit based on the GBS in order to generate the GKP qubit required for fault-tolerant QC.

\section{Conclusion} 
By virtue of many advantages compared with other physical systems, optical CVQC has attracted much interest in the last decade as a promising candidate for realizing universal and fault-tolerant QC. 
In this review, we have highlighted several topics of recent experimental and theoretical progress in terms of universality, scalability, and fault tolerance in optical CVQC. 
As we have seen, the technologies of optical multiplexing, bandwidth broadening, and integrated photonic chips pave the way towards scaling up photonic quantum computers. 
Furthermore, there are many recent theoretical efforts towards universal and fault-tolerant CVQC with bosonic codes, including reductions of the threshold for the squeezing level and proposals for the architecture of CVQC.
Among the many available bosonic codes, the GKP qubit has been shown to be the optimal code for CVQC because it not only has superior error tolerance but also requires only Gaussian operations for universality. Thus, the promising research direction towards large-scale CVQC would be to develop efficient light sources for optical GKP qubits and process these states with time-multiplexed CVQC architectures, which already have been demonstrated to perform Gaussian operations in a scalable manner.

Of course, there remains a long way towards the realization of large-scale QC. Specifically, non-Gaussian gates required for universality have not been demonstrated experimentally. In addition, the GKP qubits with a sufficient squeezing level for large-scale QC has not been generated yet in an optical setup. Furthermore, those need to be combined with the optical technologies in a scalable and low-loss configuration manner.
Nevertheless, we believe that the emerging ideas described in this review will open promising paths to these problems. These ideas will also open up new possibilities for architecture design, hardware integration, and operating software for CVQC, which finally lead to large-scale QC. We hope that this review will attract many researchers to the field of quantum information with CVs.

\section*{Acknowledgments}
This work was partly supported by JST [Moonshot R$\&$D][Grant No. JPMJMS2064], JST [Moonshot R$\&$D][Grant No. JPMJMS2061], donations from Nichia Corporation, MEXT Leading Initiative for Excellent Young Researchers, Katsu Research Encouragement Award of the University of Tokyo, Toray Science Foundation (19-6006), Research Foundation for Opto-Science and Technology, and the Kayamori Foundation of Informational Science Advancement. The authors would like to thank Yutaro Enomoto, Takahiro Kashiwazaki, Takaya Matsuura, Takahiro Mitani, Hayata Yamasaki for their helpful comments on the manuscript.

\section*{References}
\bibliography{iopart-num}

\end{document}